\documentclass[twoside,12pt,onecolumn,draftcls,conference]{IEEEtran}
\usepackage{amsmath}
\usepackage{amsfonts}
\usepackage{latexsym}
\usepackage{subfigure}
\usepackage{mathrsfs}
\def\CC{\mathbb C}
\def\EE{\mathbb E}
\def\NC{\mathcal N}

\def\SNR{\text{SNR}}

\def \PP{\mathbb P}
\def \Tr{\text{Tr}}
\newtheorem{theorem}{Theorem}[section]
\newtheorem{lemma}{Lemma}[section]

\usepackage[dvips]{graphicx}

\newcommand{\eps}{\epsilon}

\begin{document}

\title{Hierarchical Cooperation Achieves Optimal Capacity Scaling in Ad Hoc Networks}
\author{\authorblockN{Ayfer {\"O}zg{\"u}r, Olivier L\'ev\^eque}
\authorblockA{Facult\'e Informatique et Communications\\
              Ecole Polytechnique F\'ed\'erale de Lausanne\\
              1015 Lausanne, Switzerland\\
              \{ayfer.ozgur,olivier.leveque\}@epfl.ch}
\and
\authorblockN{David Tse}
\authorblockA{Wireless Foundations, Dept. of EECS\\
University of California at Berkeley\\
Berkeley, CA 94720, USA\\
dtse@eecs.berkeley.edu}} \maketitle

\begin{abstract}
$n$ source and destination pairs randomly located in an area want to
communicate with each other. Signals transmitted from one user to
another at distance $r$ apart are subject to a power loss of
$r^{-\alpha}$ as well as a random phase.
We identify the scaling laws of the information theoretic capacity
of the network. In the case of dense networks, where the area is fixed
and the density of nodes increasing, we show that the total capacity
of the network scales {\em linearly} with $n$. This improves on the
best known achievability result of $n^{2/3}$ of \cite{AS06}. In the
case of extended networks, where the density of nodes is fixed and
the area increasing linearly with $n$, we show that this capacity
scales as $n^{2-\alpha/2}$ for $2\le \alpha <3$ and $\sqrt{n}$ for
$\alpha \ge 3$. The best known earlier result \cite{XK06} identified
the  scaling law for $\alpha > 4$. Thus, much better scaling than
multihop can be achieved in dense networks, as well as in extended
networks with low attenuation. The performance gain is achieved by
intelligent node cooperation and distributed MIMO communication. The
key ingredient is a {\em hierarchical} and {\em digital}
architecture for nodal exchange of information for realizing the
cooperation.
\end{abstract}

\IEEEpeerreviewmaketitle

\section{Introduction}

The seminal paper by Gupta and Kumar \cite{GK00} initiated the study
of scaling laws in large ad-hoc wireless networks. Their
by-now-familiar model considers $n$ nodes randomly located in the
unit disk, each of which wants to communicate to a random
destination node at a rate $R(n)$ bits/second. They ask what is the
maximally achievable scaling of the total throughput $T(n) = n \, R(n)$
with the system size $n$. They showed that classical multihop
architectures with conventional single-user decoding and forwarding
of packets cannot achieve a scaling better than $O(\sqrt{n})$, and
that a scheme that uses only nearest-neighbor communication can
achieve a throughput that scales as $\Theta(\sqrt{n/\log n})$. This
gap was later closed by Franceschetti et al \cite{FDT07}, who showed
using percolation theory that the  $\Theta(\sqrt{n})$ scaling is
indeed achievable.

The Gupta-Kumar model makes certain assumptions on the
physical-layer communication technology. In particular, it assumes
that the signals received from nodes other than one particular
transmitter are interference to be regarded as noise degrading the
communication link.
Given this assumption, direct communication between source and
destination pairs  is not preferable, as the interference generated
would preclude most of the other nodes from communicating. Instead,
the optimal strategy is to confine to nearest neighbor communication
and maximize the number of simultaneous transmissions (spatial
reuse). However, this means that each packet has to be retransmitted
many times before getting to the final destination, leading to a
sub-linear scaling of system throughput.

A natural question is whether the Gupta-Kumar scaling law is a
consequence of the physical-layer technology or whether one can do
better using more sophisticated physical-layer processing. More
generally, what is the {\em information-theoretic} scaling law of ad
hoc networks? This question was first addressed by Xie and Kumar
\cite{XK04}. They showed that whenever the power path loss exponent
$\alpha$ of the environment is greater than $6$ (i.e. the received
power decays faster than $r^{-6}$ with the distance $r$ from the
transmitter), then the nearest-neighbor multihop scheme is in fact
order-optimal. They also showed that the same conclusion holds
if the power path loss is exponential in the distance $r$, a
channel model proposed recently by Franceschetti et al \cite{FBS04}.
The work \cite{XK04} was followed by several others
\cite{JVK04,LT05,XXK05,XK06,AJV06}. Successively, they improved the
threshold on the path loss exponent $\alpha$ for which multihop is
order-optimal ($\alpha > 5$ in \cite{JVK04}, $\alpha > 4.5$ in
\cite{AJV06} and $\alpha > 4$ in \cite{XK06}). However, the question
is open for the important range of $\alpha$ between $2$ and $4$,
$\alpha =2$ corresponding to free space attenuation.

There is an important difference between the network model used in
\cite{GK00} and that used in \cite{XK04} and the follow-up works.
The paper \cite{GK00} deals with {\em dense} networks, where the
total area is fixed and the density of nodes increases. The paper
\cite{XK04} and the subsequent works, on the other hand, focus on
{\em extended} networks, which scale to cover an increasing area
with the density of nodes fixed. A way to understand the difference
between the engineering implications of these two network scalings
is by drawing a parallelism with the classical notions of {\em
interference-limitedness} and {\em coverage-limitedness}, the two
operating regimes of cellular networks. Cellular networks in urban
areas tend to have dense deployments of base-stations so that
signals are received at the mobiles with sufficient signal-to-noise
ratio (SNR) but performance is limited by {\em interference} between
transmissions in adjacent cells. Cellular networks in rural areas,
on the other hand,  tend to have sparse deployments of base-stations
so that performance is mainly limited by the ability to transmit
enough power to reach all the users with sufficient signal-to-noise
ratio. Analogously, in the dense network scaling, all nodes can
communicate with each other with sufficient SNR; performance can
only be limited by interference, if at all. The Gupta-Kumar scaling
of $\sqrt{n}$ comes precisely from such interference limitation. In
the extended network scaling, the source and destination pairs are
at increasing distance from each other, and so both interference
limitation and power limitation can come into play. The network can
be either coverage-limited or interference-limited. The
information-theoretic limit on performance proved in
\cite{XK04,JVK04,LT05,XXK05,XK06,AJV06} are all based on bounding
the maximum amount of power that can be transferred across the
network and then showing that multihop achieves that bound. Hence,
what was shown by these works is that for $\alpha > 4$, when signals
attenuate fast enough, the extended network is fundamentally
coverage-limited: even with optimal cooperative relaying, the amount
of power transferred across the network cannot be larger than that
achieved by multihop.  For $\alpha$ between $2$ and $4$, when
attenuation is lower and power transfer become easier, the question
remains open whether the network is coverage-limited or
interference-limited.

Viewing the earlier results in this light, a natural first step in
completing the picture is to return to the simpler dense network as
a vehicle to focus exclusively on the issue of interference.
Can the interference limitation implied by the
Gupta-Kumar result be overcome by more sophisticated physical-layer
processing? In a recent work \cite{AS06}, Aeron and Saligrama have
showed that the answer is indeed yes: they exhibited a scheme which
yields a throughput scaling of $\Theta(n^{2/3})$ bits/second.
However, it is not clear if one can do even better. The first main
result in this paper is that, for any value of $\alpha \ge 2$,  one
can in fact achieve arbitrarily close to {\em linear} scaling: for
any $\epsilon>0$, we present a scheme that achieves an aggregate
rate of $\Theta(n^{1-\eps})$. This is a surprising result: a linear
scaling means that there is essentially {\em no}
interference-limitation; the rate for {\em each} source-destination
pair does not degrade significantly even as one puts more and more
nodes in the network. It is easy to show that one cannot get a
better capacity scaling than $O(n \log n)$, so our scheme is close
to optimal.

To achieve linear scaling, one must be able to perform {\em many}
simultaneous long-range communications. A physical-layer technique
which achieves this is MIMO (multi-input multi-output): the use of
multiple transmit and receive antennas to multiplex several streams
of data and transmit them simultaneously. MIMO was originally
developed in the point-to-point setting, where the transmit antennas
are co-located at a single transmit node, each transmitting one data
stream, and the receive antennas are co-located at a single receive
node, jointly processing the vector of received observations at the
antennas. A natural approach to apply this concept to the network
setting is to have both source nodes and destination nodes cooperate
in {\em clusters} to form distributed transmit and receive antenna
arrays respectively. In this way, mutually interfering signals can
be turned into useful ones that can be jointly decoded at the
receive cluster and spatial multiplexing gain can be realized. In
fact, if {\em all} the nodes in the network could cooperate for
free, then a classical MIMO result \cite{Fos96,Tel99} says that a
sum rate scaling proportional to $n$ could be achieved. However,
this may be over-optimistic : communication between nodes is
required to set up the cooperation and this may drastically reduce
the useful throughput. The Aeron-Saligrama scheme is MIMO-based and
its performance is precisely limited by the cooperation overhead
between receive nodes. Our main contribution is to introduce a new
{\em multi-scale, hierarchical} cooperation architecture without
significant overhead.
Such cooperation first takes place between nodes within very small
local clusters to facilitate MIMO communication over a larger
spatial scale. This can then be used as a communication
infrastructure for cooperation within larger clusters at the next
level of the hierarchy. Continuing on this fashion, cooperation can
be achieved at an almost global scale.


The result in the dense network builds the foundation for
understanding the extended network in the low-attenuation regime of
the path loss exponent $\alpha$ between $2$ and $4$. Cooperative
MIMO communication provides not only a degree of freedom gain but
also a power gain, obtained by combining signals received at the
different nodes. This power gain is not very important in the dense
setting, since there is already sufficient SNR in any direct
communication between individual nodes and the capacity is only {\em
logarithmic} in the SNR. In the extended setting, however, this
power gain becomes very important, since the power transferred
between an individual source and destination pair vanishes due to
channel attenuation. The operation is in the low SNR regime where the
capacity is {\em linear} in the SNR. Cooperation between nodes can
significantly boost up the power transfer. In fact, it can be shown
that the capacity of long range $n$ by $n$ cooperative MIMO
transmission scales exactly like the total received power. This
total received power scales like $n^{2-\alpha/2}$. We show that a
simple modification of our hierarchical cooperation scheme to the
extended setting can achieve a network total throughput arbitrarily
close to this cooperative MIMO scaling. Thus, for $\alpha <3$, our
scheme performs strictly better than multihop.

Can we do better? Recall that earlier results in
\cite{XK04,JVK04,LT05,XXK05,XK06,AJV06} are all based on upper
bounding the amount of power transferred across cutsets of the
network. It turns out that their upper bounds are tight when $\alpha
> 4$ but not tight for $\alpha$ between $2$ and $4$. By evaluating
exactly the scaling of the power transferred, we show that it
matches the performance of the hierarchical scheme for $\alpha$
between $2$ and $3$ and that of the multihop scheme for $\alpha>3$.
More precisely, we obtain the following tight characterization
for the scaling exponent for all $\alpha$ in the extended case:
$$
e(\alpha) := \lim_{n \rightarrow \infty} \frac{\log
C_n(\alpha)}{\log n} = \left \{ \begin{array}{cc} 2 -
\frac{\alpha}{2} & 2 \le \alpha  \le 3 \\ \frac{1}{2} & \alpha > 3
\end{array} \right .
$$
where $C_n(\alpha)$ is the total capacity of the network. In
particular, when $\alpha = 2$, linear capacity scaling can be
achieved, even in the extended case. Note that the capacity is
limited by the power transferred for {\em all} $\alpha \ge 2$; hence
the extended network is fundamentally {\em coverage-limited}, even
for $\alpha$ between $2$ and $4$. For $\alpha > 3$, multihop is
sufficient in transferring the optimal amount of power; for $\alpha
< 3$, when the attenuation is slower, cooperative MIMO is needed to
provide the power gain and also enough degrees of freedom to
operate in the power-efficient regime. Just like in the dense
setting, interference limitation does not play a significant role,
as far as capacity scaling is concerned. Cooperative MIMO takes care
of that.




Our approach to the problem is to first look at the dense case to
isolate the issue of interference and then to tackle the extended case.
But the dense scaling is also of interest on its own right. It is
relevant whenever one wants to design networks to serve many nodes,
all within communication range of each other (within a campus, an
urban block, etc.). This scaling is also a reasonable model to study
problems such as {\em spectrum sharing}, where many users in a
geographical area are sharing a wide band of spectrum. Consider the
scenario where we segregate the total bandwidth into many orthogonal
bands, one for each separate network supporting a {\em fixed} number
of users. As we increase the number of users, the number of such
segregated networks increases but the {\em spectral efficiency}, in
bits/s/Hz, does not scale with the {\em total} number of users. In
contrast, if we build one large ad hoc network for all the users on
the entire bandwidth, then our result says that the spectral
efficiency actually increases {\em linearly} with the number of
users. The gain is coming from a {\em network} effect via
cooperation between the many nodes in the system.

The rest of the paper is summarized as follows. In Section
\ref{sec:model}, we present the model and discuss the various
assumptions. Section \ref{sec:outline} contains the main result for
dense networks  and an outline of the proposed architecture together
with a back-of-the-envelope analysis of its performance. The details
of its performance analysis are given in Section \ref{sec:details}.
Section \ref{sec:extended} characterizes the scaling law for
extended networks. Section \ref{sec:discussions} discusses the
limitations of the model and the results.
Section \ref{sec:conclusions} contains our conclusions.

\section{Model}
\label{sec:model}

There are $n$ nodes uniformly and independently distributed in a
square of unit area in the dense scaling (Sections~\ref{sec:outline}
and \ref{sec:details}) and a square of area $n$ in the extended scaling
(Section~\ref{sec:extended}). Every node is both a source and
a destination. The sources and destinations are paired up one-to-one
in an arbitrary fashion. Each source has the same traffic rate $R(n)$ to
send to its destination node and a common average transmit power
budget of $P$ Watts. The total throughput of the system is $T(n) = n
R(n)$.\footnote{In the sequel, whenever we say a total throughput
$T(n)$ is achievable, we implicitly mean that a rate of $T(n)/n$ is
achievable for every source-destination pair.}

We assume that communication takes place over a flat channel of
bandwidth $W$ Hz around a carrier frequency of $f_c$, $f_c \gg W$.
The complex baseband-equivalent channel gain between  node $i$ and
node $k$ at time $m$ is given by:
\begin{equation}
\label{eq:ch_model} H_{ik}[m] = \sqrt{G}r_{ik}^{-\alpha/2} \exp(j
\theta_{ik}[m])
\end{equation}
where $r_{ik}$ is the distance between the nodes, $\theta_{ik}[m]$
is the random phase at time $m$, uniformly distributed in $[0,2\pi]$
and $\{\theta_{ik}[m], 1\leq i\leq n, 1\leq k\leq n\}$ is a
collection of i.i.d.~random processes. The $\theta_{ik}[m]$'s and the
$r_{ik}$'s are also assumed to be independent. The parameters $G$
and $\alpha \ge 2$ are assumed to be constants; $\alpha$ is called
the path loss exponent. For example, under free-space line-of-sight
propagation, Friis' formula applies and
\begin{equation}
\label{eq:friis} |H_{ik}[m]|^2 = \frac{G_{Tx} \cdot G_{Rx}}{\left
(4\pi r_{ik}/\lambda_c \right)^2}
\end{equation}
so that
$$ G = \frac{G_{Tx} \cdot G_{Rx} \cdot \lambda_c^2}{ 16 \pi^2}, \qquad \alpha = 2.$$
where $G_{Tx}$ and $G_{Rx}$ are the transmitter and receiver antenna
gains respectively and $\lambda_c$ is the carrier wavelength.

Note that the channel is random, depending on the location of the
users and the phases. The locations are assumed to be fixed over the
duration of the communication. The phases are assumed to vary in a
stationary ergodic manner (fast fading).\footnote{With more
technical efforts, we believe our results can be extended to the
slow fading setting where the phases are fixed as well. See the
remark at the end of Appendix~\ref{app:MIMO} for further discussion
on this point.} We assume that the channel gains are known at all
the nodes. The signal received by node $i$ at time $m$ is given by
$$
Y_i[m]=\sum_{k=1}^{n}H_{ik}[m]X_k[m]+Z_i[m]
$$
where $X_k[m]$ is the signal sent by node $k$ at time $m$ and
$Z_i[m]$ is white circularly symmetric Gaussian noise of variance
$N_0$ per symbol.

Several comments about the model are in order:

\begin{itemize}

\item The path loss model is based on a {\em far-field} assumption: the
distance $r_{ik}$ is assumed to be much larger than the carrier
wavelength. When the distance is of the order or shorter than the
carrier wavelength, the simple path loss model obviously does not
hold anymore as path loss can potentially become path ``gain". The
reason is that near-field electromagnetics now come into play.

\item The phase $\theta_{ik}[m]$ depends on the distance between the nodes
modulo the carrier wavelength \cite{TV05}. The random phase model is
thus also based on a far-field assumption: we are assuming that the
nodes' separation is at a much larger spatial scale compared to the
carrier wavelength, so that the phases can be modelled as completely
random and independent of the actual positions.

\item It is realistic to assume the variation of the phases since they
vary significantly when users move a distance of the order of the
carrier wavelength (fractions of a meter). The positions determine
the path losses and they on the other hand vary over a much larger
spatial scale. So the positions are assumed to be fixed.

\item We essentially assume a line-of-sight type environment and ignore
multipath effects. The randomness in phases is sufficient for the
long range MIMO transmissions needed in our scheme. With multipaths,
there is a further randomness due to random constructive and
destructive interference of these paths. It can be seen that our
results easily extend to the multipath case.

\end{itemize}

We will discuss further the limitations of this model in Section
\ref{sec:discussions} after we present our results.


\section{Main Result for Dense Networks}
\label{sec:outline} We first give an information-theoretic upper
bound on the achievable scaling law for the aggregate throughput in
the network. Before starting to look for good communication
strategies, Theorem~\ref{thm:thm2} establishes the best we can hope
for.
\smallskip
\smallskip
\begin{theorem}
\label{thm:thm2} The aggregate throughput in the network with $n$
nodes is bounded above by
$$
T(n)\leq K^{\prime}n\log{n}
$$
with high probability\footnote{i.e.~probability going to $1$ as
system size grows.} for some constant $K^{\prime}>0$ independent of
$n$.
\end{theorem}
\smallskip
\smallskip

\noindent\textit{Proof:} Consider a  source-destination pair $(s,d)$
in the network. The transmission rate $R(n)$ from source node $s$ to
destination node $d$ is upper bounded by the capacity of the
single-input multiple-output (SIMO) channel between source node $s$
and the rest of the network. Using a standard formula for this
channel (see eg. \cite{TV05}), we get:
$$
R(n)\leq\log\left(1+\frac{P}{N_0}\sum_{\substack{i=1 \\ i\neq
s}}^{n}|H_{is}|^2\right)=\log\left(1+\frac{P}{N_0}\sum_{\substack{i=1
\\ i\neq s}}^{n}\frac{G}{r_{is}^{\alpha}}\right).
$$
It is easy to see that in a random network with $n$ nodes uniformly
distributed on a fixed two-dimensional area, the minimum distance
between any two nodes in the network is larger than
$\frac{1}{n^{1+\delta}}$ with high probability, for any $\delta>0$.
Consider one specific node in the network which is at distance
larger than $\frac{1}{n^{1+\delta}}$ to all other nodes in the
network. This is equivalent to saying that there are no other nodes
inside a circle of area $\frac{\pi}{n^{2+2\delta}}$ around this
node. The probability of such an event is
$\left(1-\frac{\pi}{n^{2+2\delta}} \right)^{n-1}$. Moreover, the
minimum distance between any two nodes in the network is larger than
$\frac{1}{n^{1+\delta}}$ only if this condition is satisfied for all
nodes in the network. Thus by the union bound we have,
\begin{align*}
P \left(\textrm{minimum distance in the network is smaller than
$\frac{1}{n^{1+\delta}}$} \right)&\leq
n\left(1-\left(1-\frac{\pi}{n^{2+2\delta}} \right)^{n-1}\right)
\end{align*}
which decreases to zero as $1/n^{2\delta}$ with increasing $n$.

Hence using this fact on the minimum distance in the network, we
obtain
$$
R(n)\leq \log \left( 1+\frac{GP}{N_0}n^{\alpha(1+\delta)+1}
\right)\leq K^{\prime}\log n
$$
for some constant $K^{\prime}>0$ independent of $n$ for all-source
destination pairs in the network with high probability. The theorem
follows.\hfill $\square$
\smallskip
\smallskip

In the view of what is ultimately possible, established by
Theorem~\ref{thm:thm2}, we are now ready to state the main result of
this paper.
\smallskip
\smallskip

\begin{theorem}
\label{thm:thm1} Let $\alpha\geq 2$. For any $\epsilon>0$,
there exists a constant $K_\epsilon>0$ independent of $n$
such that with high probability, an aggregate throughput
$$
T(n) \geq K_\epsilon \, n^{1-\epsilon}
$$
is achievable in the network for all possible pairings between
sources and destinations.
\end{theorem}
\smallskip
\smallskip

Theorem~\ref{thm:thm1} states  that it is actually possible to
perform arbitrarily close to the bound given in
Theorem~\ref{thm:thm2}.
The two theorems together establish the capacity scaling for the
network up to logarithmic terms. Note how dramatically different is
this new linear capacity scaling law  from the well-known throughput
scaling of $\Theta{(\sqrt{n})}$ implied by \cite{GK00,FDT07} for the
same model. Note also that the upper bound in Theorem~\ref{thm:thm2}
assumes a genie-aided removal of interference between simultaneous
transmissions from different sources. By proving
Theorem~\ref{thm:thm1}, we will show that it is possible to mitigate
such interference without a genie but with cooperation between the
nodes.

The proof of Theorem~\ref{thm:thm1} relies on the construction of an
explicit scheme that realizes the promised scaling law. The
construction is based on recursively using the following key lemma,
which addresses the case when $\alpha > 2$.

\smallskip
\smallskip
\begin{lemma}
\label{lem:lem1} Consider $\alpha >2$ and a network with $n$ nodes
subject to interference from external sources. The signal received
by node $i$ is given by
$$
Y_i=\sum_{k=1}^{n}H_{ik}X_k+Z_i+I_i
$$
where $I_i$ is the external interference signal received by node
$i$. Assume that $\{I_i, 1\leq i\leq n\}$ is a collection of
uncorrelated zero-mean stationary and ergodic random processes with
power $P_{I_i}$ upper bounded by
$$
P_{I_i}\leq K_I,\qquad1\leq i\leq n
$$
for a constant $K_I > 0$ independent of $n$. Let us assume there
exists a scheme such that for each $n$, with probability at least
$1-e^{-n^{c_1}}$ achieves an aggregate throughput
$$
T(n)\geq K_1 \, n^{b}
$$
for every possible source-destination pairing in this network of $n$
nodes. $K_1$ and $c_1$ are positive constants independent of $n$ and
the source-destination pairing, and $0 \le b<1$. Let us also assume
that the per node average power budget required to realize this
scheme is upper bounded by $P/n$ as opposed to $P$.

Then one can construct  another scheme for this network that
achieves a {\em higher} aggregate throughput
$$
T(n)\geq K_2 \, n^{\frac{1}{2-b}}
$$
for every source-destination pairing in the network, where $K_2>0$
is another constant independent of $n$ and the pairing.  Moreover,
the failure rate for the new scheme is upper bounded by
$e^{-n^{c_2}}$ for another positive constant $c_2$ while the per
node average power needed to realize the scheme is also upper
bounded by $P/n$.
\end{lemma}
\smallskip
\smallskip

Lemma~\ref{lem:lem1} is the key step to build a hierarchical
architecture. Since $\frac{1}{2-b} > b$ for $0\le b < 1$, the new
scheme is always better than the old. We will now give a rough
description of how the new scheme can be constructed given the old
scheme, as well as a back-of-the-envelope analysis of the scaling
law it achieves. Next section is devoted to its precise description
and performance analysis.

The constructed scheme is based on clustering and long-range MIMO
transmissions between clusters. We divide the network into clusters
of $M$ nodes.
Let us focus for now on a particular source node $s$ and its
destination node $d$. $s$ will send  $M$ bits to $d$ in 3 steps:

\begin{itemize}

\item [(1)] Node $s$ will distribute its $M$ bits among the $M$ nodes in its cluster, one for each node;

\item [(2)] These nodes together can then form a distributed transmit antenna array, sending the $M$ bits {\em simultaneously} to the destination cluster where $d$ lies;

\item [(3)] Each node in the destination cluster obtained one observation from the MIMO transmission, and it
 quantizes and ships the observation back to $d$,
 which can then do joint MIMO processing of all the observations and decode the $M$ transmitted bits.

\end{itemize}

From the network point of view, all source-destination pairs have to
eventually accomplish these three steps. Step 2 is long-range
communication and only one source-destination pair can operate at
the same time. Steps 1 and 3 involve local communication and can be
parallelized across source-destination pairs. Combining all this
leads to three phases in the operation of the network:

\textbf{Phase 1: Setting Up Transmit Cooperation} Clusters work in
parallel. Within a cluster, each source node has to distribute $M$
bits to the other nodes, 1 bit for each node, such that at the end
of the phase each node has 1 bit from each of the source nodes in
the same cluster. Since there are $M$ source nodes in each cluster,
this gives a traffic demand of exchanging $M^2$ bits. (Recall our
assumption that each node is a source for some communication request
and destination for another.) The key observation is that this is
similar to the original problem of communicating between $n$ source
and destination pairs, but on a network of size $M$. More
specifically, this traffic demand of exchanging $M^2$ bits is
handled by setting up $M$ sub-phases, and assigning $M$
source-destination pairs for each sub-phase. Since our channel model
is scale invariant, note that the scheme given in the hypothesis of
the lemma can be used in each sub-phase by simply scaling down the
power with cluster area. Having aggregate throughput $M^b$, each
sub-phase is completed in $M^{1-b}$ time slots while the whole phase
takes $M^{2-b}$ time slots. See Figure 1.
\begin{figure}
\begin{center}
\input{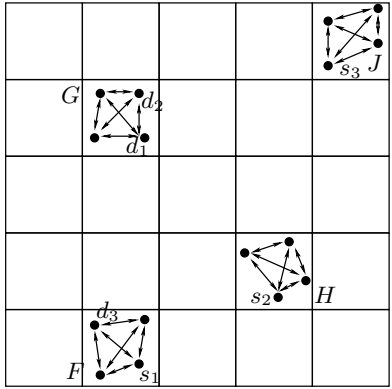}
\label{fig:phase1_sec3} \caption{Nodes inside clusters $F$, $G$, $H$
and $J$ are illustrated while exchanging bits in Phases 1 and 3.
Note that in Phase 1 the exchanged bits are the source bits whereas
in Phase 3 they are the quantized MIMO observations. Clusters work
in parallel. In this and the following figure Fig.~2, we highlight
three source-destination pairs $s_1-d_1$, $s_2-d_2$ and $s_3-d_3$,
such that nodes $s_1$ and $d_3$ are located in $F$, nodes $s_2$ and
$s_3$ are located in $H$ and $J$ respectively, and nodes $d_1$ and
$d_2$ are located in $G$.}
\end{center}
\end{figure}

\textbf{Phase 2: MIMO Transmissions} We perform successive
long-distance MIMO transmissions between source-destination pairs,
one at a time. In each one of the MIMO transmissions , say one
between $s$ and $d$, the $M$ bits of $s$  are simultaneously
transmitted by the $M$ nodes in its cluster to the $M$ nodes in the
cluster of $d$. Each of the long-distance MIMO transmissions are
repeated for each source node in the network, hence we need $n$ time
slots to complete the phase. See Figure 2.

\begin{figure*}[tbp]
\begin{center}
\label{fig:phase2}
\input{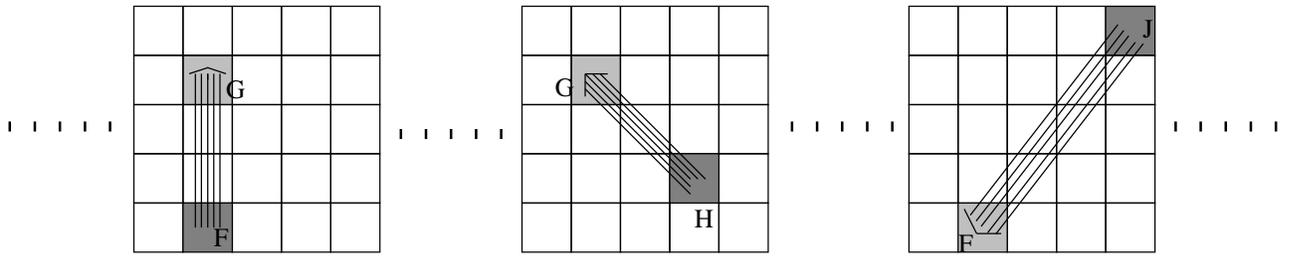}
\end{center}
\caption{Successive MIMO transmissions are performed between
clusters. The first figure depicts MIMO transmission from cluster
$F$ to $G$, where bits originally belonging to $s_1$ are
simultaneously transmitted by all nodes in $F$ to all nodes in $G$.
The second MIMO transmission is from $H$ to $G$, while now bits of
source node $s_2$ are transmitted by nodes in $H$ to nodes in $G$.
The third picture illustrates MIMO transmission from cluster $J$ to
$F$.}
\end{figure*}

\textbf{Phase 3: Cooperate to Decode} Clusters work in parallel.
Since there are $M$ destination nodes inside the clusters, each
cluster received $M$ MIMO transmissions in phase 2, one intended for
each of the destination nodes in the cluster. Thus, each node in the
cluster has  $M$ received observations, one  from each of the MIMO
transmissions, and each observation is to be conveyed to a different
node in its cluster. Nodes quantize each observation into fixed $Q$
bits so there are now a total of at most $QM^2$ bits to exchange
inside each cluster. Using exactly the same scheme as in Phase 1, we
conclude the phase in $QM^{2-b}$ time slots. See again Figure 1.

Assuming that each destination node is able to decode the
transmitted bits from its source node from the $M$ quantized signals
it gathers by the end of Phase 3, we can calculate the rate of the
scheme as follows: Each source node is able to transmit $M$ bits to
its destination node, hence $nM$ bits in total are delivered to
their destinations in $M^{2-b}+n+QM^{2-b}$ time slots, yielding an
aggregate throughput of $$ \frac{nM}{M^{2-b}+n+ QM^{2-b}}
$$ bits per time slot. Maximizing this throughput by choosing
$M^{\frac{1}{2-b}}$ yields $T(n)=\frac{1}{2+Q}n^{\frac{1}{2-b}}$
for the aggregate throughput which is the result in
Lemma~\ref{lem:lem1}.

Clusters can work in parallel in phases 1 and 3 because for $\alpha
> 2$, the aggregate interference at a particular cluster caused by
other active nodes is bounded, moreover the interference signals
received by different nodes in the cluster are zero-mean and
uncorrelated satisfying the assumptions of Lemma~\ref{lem:lem1}. For
$\alpha=2$, the aggregate interference scales like $\log n$, leading
to a slightly different version of the lemma.

\smallskip
\smallskip
\begin{lemma}
\label{lem:lem2} Consider $\alpha=2$ and a network with $n$ nodes
subject to interference from external sources. The signal received
by node $i$ is given by
$$
Y_i=\sum_{k=1}^{n}H_{ik}X_k+Z_i+I_i
$$
where $I_i$ is the external interference signal received by node
$i$. Assume that $\{I_i, 1\leq i\leq n\}$ is a collection of
uncorrelated zero-mean stationary and ergodic random processes with
power $P_{I_i}$ upper bounded by
$$
P_{I_i}\leq K_I \log n,\qquad 1\leq i\leq n
$$
for a constant $K_I\geq 0$ and independent of $n$. Let us assume
there exists a scheme such that for each $n$ with failure
probability at most $e^{-n^{c_1}}$, achieves an aggregate throughput
$$
T(n)\geq K_1\frac{n^{b}}{\log n}
$$
for every source-destination pairing in this network. $K_1$ and
$c_1$ are positive constants independent of $n$ and the
source-destination pairing, and $0 \leq b<1$. Let us also assume
that the average power budget required to realize this scheme is
upper bounded by $P/n$, as opposed to $P$

Then one can construct  another scheme for this network that
achieves a {\em higher} aggregate throughput scaling
$$
T(n)\geq K_2\frac{n^{\frac{1}{2-b}}}{(\log n)^2}
$$
for every source-destination pairing, where $K_2>0$ is another
constant independent of $n$ and the pairing. Moreover, the failure
rate for the new scheme is upper bounded by $e^{-n^{c_2}}$ for
another positive constant $c_2$ while the per node average power
needed to realize the scheme is also upper bounded by $P/n$.
\end{lemma}
\smallskip
\smallskip

We can now use Lemma~\ref{lem:lem1} and \ref{lem:lem2} to prove
Theorem~\ref{thm:thm1}.
\smallskip
\smallskip

 \noindent\textit{Proof of
Theorem~\ref{thm:thm1}:} We only focus on the case of $\alpha >2$.
The case of $\alpha =2$ proceeds similarly.

We start by observing that the simple scheme of transmitting
directly between the source-destination pairs one at a time (TDMA)
satisfies the requirements of the lemma. The aggregate throughput is
$\Theta(1)$, so $b=0$. The failure probability is $0$. Since each
source is only transmitting $\frac{1}{n}$th of the time and the
distance between the source and its destination is bounded, the
average power consumed per node is of the order of $\frac{1}{n}$.

As soon as we have a scheme to start with, Lemma~\ref{lem:lem1} can
be applied recursively, yielding a scheme that achieves higher
throughput at each step of the recursion. More precisely, starting
with a TDMA scheme with $b=0$ and applying Lemma~\ref{lem:lem1}
recursively $h$ times, one gets a scheme achieving
$\Theta(n^{\frac{h}{h+1}})$ aggregate throughput. Given any
$\eps>0$,  we can now choose $h$ such that $\frac{h}{h+1} \geq
1-\eps$ and we get a scheme that achieves $\Theta(n^{1-\epsilon})$
aggregate throughput scaling with high probability. This concludes
the proof of Theorem~\ref{thm:thm1}.\hfill $\square$
\smallskip
\smallskip

Gathering everything together, we have built a hierarchical scheme
to achieve the desired throughput. At the lowest level of the
hierarchy, we use the simple TDMA scheme to exchange bits for
cooperation among small clusters. Combining this with longer range
MIMO transmissions, we get a higher throughput scheme for
cooperation among nodes in larger clusters at the next level of the
hierarchy. Finally, at the top level of the hierarchy, the
cooperation clusters are almost the size of the network and the MIMO
transmissions are over the global scale to meet the desired traffic
demands. Figure 3 shows the resulting hierarchical scheme with a
focus on the top two levels.

\begin{figure*}[tbp]
\begin{center}
\label{fig:hier}
\input{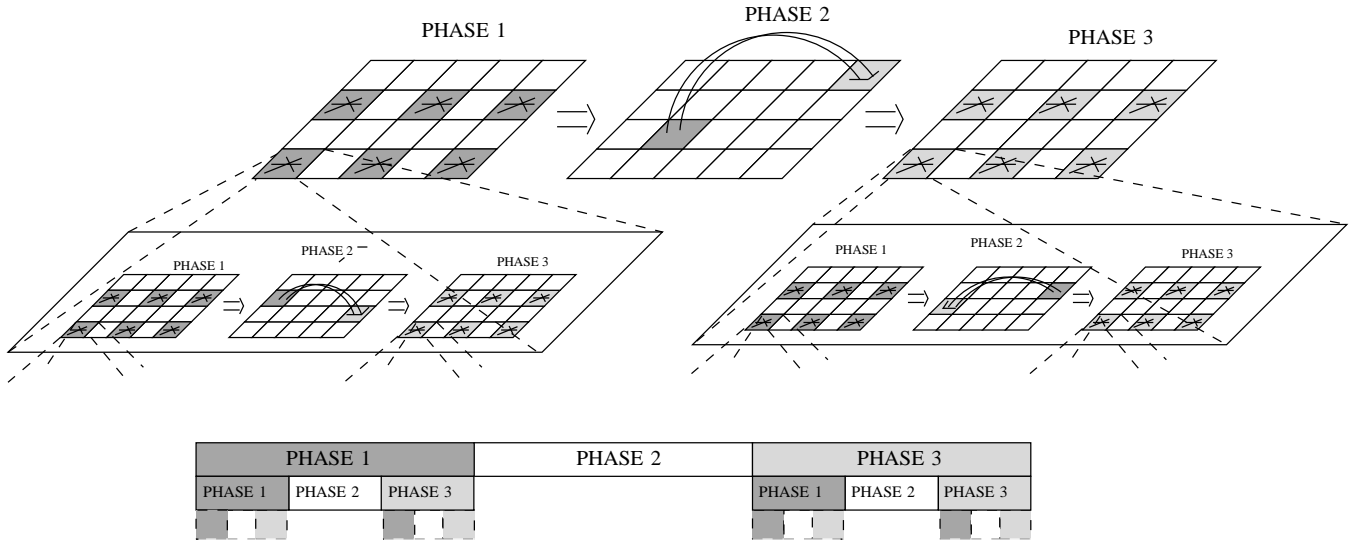}
\end{center}
\caption{The time division in a hierarchical scheme as well as the
salient features of the three phases are illustrated.}
\end{figure*}

It is important to understand the aspects of the channel model which
the scheme made use of in achieving the linear capacity scaling:

\begin{itemize}

\item the random channel phases enable the long-range MIMO
transmissions.

\item the path attenuation decay law $1/r^\alpha$ ($\alpha \ge 2$)
ensures that the {\em aggregate} signals from far away nodes are
much weaker than signals from close-by nodes. This enables spatial
reuse.

\end{itemize}

Note that the second property is exactly the same one which allows
multi-hop schemes to achieve the $\sqrt{n}$-scaling in the paper by
Gupta-Kumar \cite{GK00} and in many others after that. Although the
gain between nearby nodes becomes unbounded as $n \rightarrow
\infty$ in the model, the received signal-to-interference-plus-noise
ratio (SINR) is always bounded in the scheme at all levels of the
hierarchy. The scheme does not communicate with unbounded SINR,
although it is possible in the model.

\section{Detailed Description and Performance Analysis}
\label{sec:details}

In this section, we concentrate in  more detail on the scheme that
proves Lemma~\ref{lem:lem1} and Lemma~\ref{lem:lem2}. We first focus
on Lemma~\ref{lem:lem1} and then extend the proof to
Lemma~\ref{lem:lem2}.  As we have already seen in the previous
section, we start by dividing the unit square into smaller squares
of area $A_c=\frac{M}{n}$. Since the node density is $n$, there will
be on average $M$ nodes inside each of these small squares. The
following lemma upper bounds the probability of having large
deviations from the average.
\smallskip
\smallskip
\begin{lemma}
\label{lem:sec4_1} Let us partition a unit area network of size $n$
into cells of area $A_c$, where $A_c$ can be a function of $n$. The
number of nodes inside each cell is between
$\left((1-\delta)A_cn,(1+\delta)A_cn\right)$ with probability larger
than $1-\frac{1}{A_c}e^{-\Lambda(\delta)A_cn}$ where
$\Lambda(\delta)$ is independent of $n$ and satisfies
$\Lambda(\delta)>0$ when $\delta>0$.
\end{lemma}
\smallskip
\smallskip

Applying Lemma~\ref{lem:sec4_1}  to the squares of area $M/n$, we
see that all squares contain order $M$ nodes with probability larger
than $1-\frac{n}{M}e^{-\Lambda(\delta)M}$. We assume $M^{\gamma}$
where $0<\gamma\leq 1$ in which case this probability tends to $1$
as $n$ increases. In the following discussion, we will need a
stronger result, namely each of the $8$ possible halves of a square
should contain order $M/2$ nodes with high probability which again
follows from the lemma together with the union bound. This condition
is sufficient for our below analysis on scaling laws to hold.
However, in order to simplify the presentation, we assume that there
are exactly $M/2$ nodes inside each half, thus exactly $M$ nodes in
each square. The clustering is used to realize a distributed MIMO
system in three successive steps:

\textbf{Phase 1: Setting Up Transmit Cooperation} In this phase,
source nodes distribute their data streams over their clusters and
set up the stage for the long-range MIMO transmissions that we want
to perform in the next phase. Clusters work in parallel according to
the $9$-TDMA scheme depicted in Figure~4, which divides the total
time for this phase into $9$ time-slots and assigns simultaneous
operation to clusters that are sufficiently separated. Let us focus
on one specific source node $s$ located in cluster $S$ with
destination node $d$ in cluster $D$. Node $s$ will divide a block of
length $LM$ bits of its data stream into $M$ sub-blocks each of
length $L$ bits, where $L$ can be arbitrarily large but bounded. The
destination of each sub-block in Phase 1 depends on the relative
position of clusters $S$ and $D$:
\begin{enumerate}
\item[(1)] If $S$ and $D$ are either the same cluster or are not neighboring clusters: One sub-block is to be kept in $s$ and the rest $M-1$ sub-blocks are to transmitted to the other $M-1$ nodes located in $S$, one sub-block for each node.
\item[(2)] If $S$ and $D$ are neighboring clusters: Divide the cluster $S$ into two halves, each of area $A_c/2$, one half located close to the border with $D$ and the second half located farther to $D$. The $M$ sub-blocks of source node $s$ are to be distributed to the $M/2$ nodes located in the second half cluster (farther to $D$), each node gets two sub-blocks.
\end{enumerate}

Since the above traffic is required for every source node in cluster
$S$, we end up with a highly uniform traffic demand of delivering
$M\times LM$ bits in total to their destinations. A key observation
is that the problem can be separated into sub-problems, each similar
to our original problem, but on a network size $M$ and area $A_c$.
More specifically, the  traffic of transporting $LM^2$ bits can be
handled by organizing $M$ sessions and assigning $M$
source-destination pairs for each session. (Note that due to the
non-uniformity arising from point (2) above, one might be able to
assign only $M/2$ source-destination pairs in a session and hence
need to handle the traffic demand of transporting $LM^2$ bits by
organizing up to $2M$ sessions in the extreme case instead of $M$.)
The assigned source-destination pairs in each session can then
communicate $L$ bits. Since our channel model is scale invariant,
the scheme in the hypothesis of Lemma~\ref{lem:lem1} can be used to
handle the traffic in each session, by simply scaling down the
powers of the nodes by $(A_c)^{\alpha/2}$. Hence, the power used by
each node will be bounded above by $\frac{P(A_c)^{\alpha/2}}{M}$.
The scheme is to be operated simultaneously inside all the clusters
in the $9$-TDMA scheme, so we need to ensure that the resultant
inter-cluster interference satisfies the properties in
Lemma~\ref{lem:lem1}.
\smallskip
\smallskip
\begin{lemma}
\label{lem:sec4_2} Consider clusters of size $M$ and area $A_c$
operating according to 9-TDMA scheme in Figure~4 in a network of
size $n$. Let each node be constrained to an average power
$\frac{P(A_c)^{\alpha/2}}{M}$. For $\alpha>2$, the interference
power received by a node from the simultaneously operating clusters
is upper bounded by a constant $K_{I_1}$ independent of $n$.  For
$\alpha=2$, the interference power is bounded by $K_{I_2}\log n$ for
$K_{I_2}$ independent of $n$. Moreover, the interference signals
received by different nodes in the cluster are zero-mean and
uncorrelated.
\end{lemma}
\smallskip
\smallskip

Let us for now concentrate on the case $\alpha>2$. By
Lemma~\ref{lem:sec4_2}, the inter-cluster interference has bounded
power and is uncorrelated across different nodes. Thus, the strategy
in the hypothesis of Lemma~\ref{lem:lem1} can achieve an aggregate
rate $K_1 M^{b}$ in each session for some $K_1>0$, with probability
larger than $1-e^{-M^{c_1}}$. Using the union bound, with
probability larger than $1-2ne^{-M^{c_1}}$, the aggregate rate $K_1
M^{b}$ is achieved inside all sessions in all clusters in the
network. (Recall that the number of sessions in one cluster can be
$2M$ in the extreme case and there are $n/M$ clusters in total.)
With this aggregate rate, each session can be completed in at most
$(L/K_1)M^{1-b}$ channel uses and $2M$ successive sessions are
completed in $(2L/K_1)M^{2-b}$ channel uses. Using the 9-TDMA
scheme, the phase is completed in less than $(18L/K_1)M^{2-b}$
channel uses all over the network with probability larger than
$1-2ne^{-M^{c_1}}$.

\begin{figure*}[tbp]
\begin{center}
\label{fig: phase1}
\input{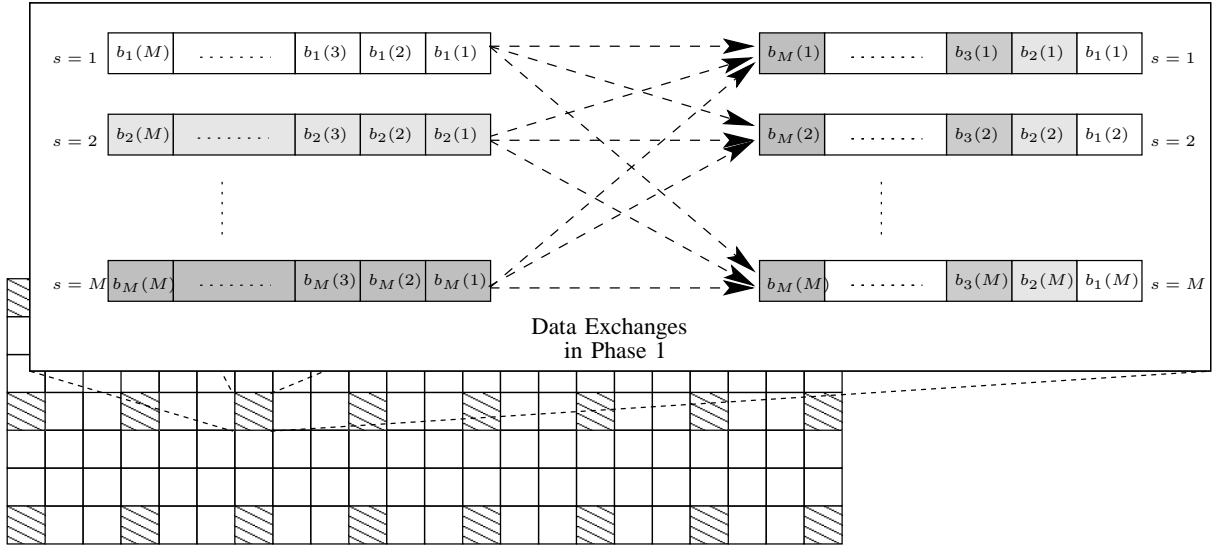}
\end{center}
\caption{Buffers of the  nodes in a cluster are illustrated before
and after the data exchanges in Phase 1. The data stream of the
source nodes are distributed to the $M$ nodes in the network as
depicted. $b_s(j)$ denotes the $j$'th sub-block of the source node
$s$. Note the 9-TDMA scheme that is employed over the network in
this phase.}
\end{figure*}

\textbf{Phase 2: MIMO Transmissions} In this phase, we are
performing the actual MIMO transmissions for all the
source-destination pairs serially, i.e. one at a time. A MIMO
transmission from source $s$ to destination $d$ involves the $M$ (or
$M/2$) nodes in the cluster $S$, where $s$ is in (referred to as the
source cluster for this MIMO transmission) to the $M$ (or $M/2$)
nodes of the cluster $D$, where $d$ is in (referred to as the
destination cluster of the MIMO transmission).

Let the distance between the mid-points of the two clusters be
$r_{SD}$. If $S$ and $D$ are the same cluster, we skip the step for
this source-destination pair $s-d$. Otherwise, we operate in two
slightly different modes depending on the relative positions of $S$
and $D$ Each mode is a continuation of the operations performed in
the first phase. First consider the case where $S$ and $D$ are not
neighboring clusters. In this case, the $M$ nodes in cluster $S$
independently encode the $L$ bits-long sub-blocks they possess,
originally belonging to node $s$, into $C$ symbols by using a
randomly generated Gaussian code $\mathcal{C}$ that respects an
average transmit power constraint $\frac{P(r_{SD})^{\alpha}}{M}$.
The nodes then transmit their encoded sequences of length $C$
symbols simultaneously to the $M$ nodes in cluster $D$. The nodes in
cluster $D$ properly sample the signals they observe during the $C$
transmissions and store these samples (that we will simply refer to
as \textit{observations} in the following text), without trying to
decode the transmitted symbols. In the case where $S$ and $D$ are
neighbors, the strategy is slightly modified so that the MIMO
transmission is from the $M/2$ nodes in $S$, that possess the
sub-blocks of $s$ after Phase 1, to the $M/2$ nodes in $D$ that are
located in the farther half of the cluster to $S$. Each of these
$M/2$ nodes in $S$ possess two sub-blocks that come from $s$. They
encode each sub-block into $C$ symbols by again using a Gaussian
code of power $\frac{P(r_{SD})^{\alpha}}{M}$. The nodes then
transmit the $2C$ symbols to the $M/2$ nodes in $D$ that in turn
sample their received signals and store the observations. The
observations accumulated at various nodes in $D$ at the end of this
step are to be conveyed to node $d$ during the third phase.

After concluding the step for the pair $s-d$, the phase continues by
repeating the same step for the next source node $s+1$ in $S$ and
its destination $d^\prime$. Note that the destination cluster for
this new MIMO transmission is, in general, a different cluster
$D^\prime$, which is the one that contains the destination node
$d^\prime$. The MIMO transmissions are repeated until the data
originated from all source nodes in the network are transmitted to
their respective destination clusters. Since the step for one
source-destination pair takes either $C$ or $2C$ channel uses,
completing the operation for all $n$ source nodes in the network
requires at most $2C\times n=2Cn$ channel uses.

\textbf{Phase 3: Cooperate to Decode} In this phase, we aim to
provide each destination node, the observations of the symbols that
have been originally intended for it. With the MIMO transmissions in
the second phase, these observations have been accumulated at the
nodes of its cluster. As before, let us focus on a specific
destination node $d$ located in cluster $D$. Note that depending on
whether the source node of $d$ is located in a neighboring cluster
or not, either each of the $M$ nodes in $D$ have $C$ observations
intended for $d$, or $M/2$ of the nodes have $2C$ observations each.
Note that these observations are some real numbers that need to be
quantized and encoded into bits before being transmitted. Let us
assume that we are encoding each block of $C$ observations into $CQ$
bits, by using fixed $Q$ bits per observation on the average. The
situation is symmetric for all $M$ destination nodes in $D$, since
the cluster received $M$ MIMO transmissions in the previous phase,
one for each destination node. (The destination nodes that have
source nodes in $D$ are exception. Recall from Phase 1 and Phase 2
that in this case, each node in $D$ possesses sub-blocks of the
original data stream for the destination node, not MIMO
observations. We will ignore this case by simply assuming $L\leq CQ$
in the below computation.) The arising traffic demand of
transporting $M\times CQM$ bits in total is similar to Phase 1 and
can be handled by using exactly the same scheme in less than
$(2CQ/K_1)M^{2-b}$ channel uses. Recalling the discussion on the
first phase, we conclude that the phase can be completed in less
than $(18CQ/K_1)M^{2-b}$ channel uses all over the network with
probability larger than $1-2ne^{-M^{c_1}}$.

Note that if it were possible to encode each observation into fixed
$Q$ bits without introducing any distortion, which is obviously not
the case, the following lemma on MIMO capacity would suggest that
with the Gaussian code $\mathcal{C}$ used in Phase 2 satisfying
$L/C\geq \kappa$ for some constant $\kappa>0$, the transmitted bits
could be recovered by an arbitrarily small probability of error from
the observations gathered by the destination nodes at the end of
Phase 3.

\smallskip
\smallskip
\begin{lemma}\label{lem:sec4_3}
The mutual information achieved by the $M \times M$ MIMO
transmission between any two clusters grows at least linearly with
$M$.
\end{lemma}
\smallskip
\smallskip

The following lemma states that there is actually a way to encode
the observations using fixed number of bits per observation and at
the same time, not to degrade the performance of the overall channel
significantly, that is, to still get a linear capacity growth for
the resulting {\em quantized MIMO} channel.

\smallskip
\smallskip
\begin{lemma}\label{lem:sec4_4}
There exists a strategy to encode the observations at a fixed rate
$Q$ bits per observation and get a linear growth of the mutual
information for the resultant $M \times M$ quantized MIMO channel.
\end{lemma}
\smallskip
\smallskip

We leave the proof of the lemma to Appendix~\ref{app:quant_MIMO}
however the following small lemma may provide motivation for the
stated result. Lemma~\ref{lem:sec4_5} points out a key observation
on the way we choose our transmit powers in the MIMO phase. It is
central to the proof of Lemma~\ref{lem:sec4_4} and states that the
observations have bounded power, that does not scale with $M$. This
in turn suggests that one can use a fixed number of bits to encode
them without degrading the scaling performance of the scheme.

\smallskip
\smallskip
\begin{lemma}\label{lem:sec4_5}
In Phase 2, the power received by each node in the destination
cluster is bounded below and above by constants $P_1$ and $P_2$
respectively that are independent of $M$.
\end{lemma}
\smallskip
\smallskip

Putting it together, we have seen that the three phases described
effectively realize virtual MIMO channels achieving spatial
multiplexing gain $M$ between the source and destination nodes in
the network. Using these virtual MIMO channels, each source is able
to transmit $ML$ bits in
\begin{align*}
T_t &=T(\textrm{phase 1})+T(\textrm{phase 2})+T(\textrm{phase 3})\\
&=\frac{18L}{K_1}M^{2-b}+2Cn+\frac{18CQ}{K_1}M^{2-b}
\end{align*}
total channel uses where $L/C\geq \kappa$ for some $\kappa>0$
independent of $M$ (or $n$). This gives an aggregate throughput of
\begin{align}
T(n) &=\frac{nML}{(18L/K_1)M^{2-b}+2Cn+(18CQ/K_1)M^{2-b}} \nonumber\\
&\geq K_2n^{\frac{1}{2-b}}\label{eq:scaling_law}
\end{align}
for some $K_2>0$ independent of $n$, by choosing
$M^{\frac{1}{2-b}}$ with $0 \leq b<1$, which is the optimal choice
for the cluster size as a function of $b$. A failure arises if there
are not order $M/2$ nodes in each half cluster or the scheme used in
Phases 1 and 3 fails to achieve the promised throughput. Combining
the result of Lemma~\ref{lem:sec4_1} with the computed failure
probabilities for Phases 1 and 3 yields
$$
P_f\leq 4ne^{-M^{c_1}}+\frac{8n}{M}e^{-\Lambda(\delta)M/2}\leq
e^{-n^{c_2}}
$$
for some $c_2>0$.

Next, we show that per node average power used by the new scheme is
also bounded above by $P/n$: for Phases 1 and 3, we know that the
scheme employed inside the clusters uses average per node power
bounded above by $P A_c^{\alpha/2}/M$. Indeed, $A_c = M / n$, and
for $\alpha\geq 2$ we have
$$
\frac{P A_c^{\alpha/2}}{M}= \frac{P}{M} \, \left(\frac{M}{n}
\right)^{\alpha/2} = \frac{P}{n} \, \left( \frac{M}{n}
\right)^{\alpha/2-1} \leq \frac{P}{n}.
$$
In Phase 2, each node is transmitting with power
$\frac{P(r_{SD})^{\alpha}}{M}$ in at most fraction $M/n$  of the
total duration of the phase, while keeping silent during the rest of
the time. This yields a per node average power
$\frac{P(r_{SD})^{\alpha}}{n}$. Recall that $r_{SD}$ is the distance
between the mid-points of the source and destination clusters and
$r_{SD}<1$, which yields the upper bound $P/n$ on the per-node
average power also for the second phase.

In order to conclude the proof of Lemma~\ref{lem:lem1}, we should
note that the new scheme achieves the same aggregate throughput
scaling when the network experiences interference from the exterior.
In phases 1 and 3, this external interference with bounded power
will simply add to the inter-cluster interference experienced by the
nodes. For the MIMO phase, this will result in uncorrelated
background-noise-plus-interference at the receiving nodes which is
not necessarily Gaussian. In Appendix~\ref{app:MIMO} and
\ref{app:quant_MIMO} we prove the results stated in
Lemma~\ref{lem:sec4_3} and Lemma~\ref{lem:sec4_4} for this more
general case. This concludes the proof of
Lemma~\ref{lem:lem1}.\hfill $\square$

\smallskip
\smallskip
\noindent\textit{Proof of Lemma~\ref{lem:lem2}:} The scheme that
proves Lemma~\ref{lem:lem2} is completely similar to the one
described above. Lemma~\ref{lem:sec4_2} states that when $\alpha=2$,
the inter-cluster interference power experienced during Phases 1 and
3 is upperbounded by $K_{I_2}\log n=K_{I_2}^\prime\log M$. From the
assumptions in the lemma, there is furthermore the external
interference with power bounded by $K_I\log n$ that is adding to the
inter-cluster interference. Under these conditions, the scheme in
the hypothesis of Lemma~\ref{lem:lem2} achieves an aggregate rate
$K_1\frac{M^{b}}{\log M}$ when used to handle the traffic in these
phases. For the second phase we have the following lemma which
provides a lower bound on the spatial multiplexing gain of the
quantized MIMO channel under the interference experienced.
\smallskip
\begin{lemma}\label{lem:sec4_6}
Let the MIMO signal received by the nodes in the destination cluster
be corrupted by an interference of power $K_I\log M$, uncorrelated
over different nodes and independent of the transmitted signals.
There exists a strategy to encode these corrupted observations at a
fixed rate $Q$ bits per observation and get a $M/\log M$ growth of
the mutual information for the resulting $M \times M$ quantized MIMO
channel.
\end{lemma}
\smallskip
\smallskip
A capacity of $M/\log M$ for the resulting MIMO channel implies that there exists a code $\mathcal{C}$ that encodes $L$ bits-long sub-blocks into $C\log M$ symbols, where $L/C\geq\kappa^{\prime}$ for a constant $\kappa^{\prime}>0$, so that the transmitted bits can be decoded at the destination nodes with arbitrarily small probability of error for $L$ and $C$ sufficiently large. Hence, starting again with a block of $LM$ bits in each source node, the $LM^2$ bits in the first phase can be delivered in $(L/K_1)M^{2-b}\log M$ channel uses. In the second phase, the $L$ bits-long sub-blocks now need to be encoded into $C\log M$ symbols, hence the transmission for each source-destination pair takes $C\log M$ channel uses, the whole phase taking $Cn\log M$ channel uses. Note that there are now $CM^2\log M$ observations encoded into $CQM^2\log M$ bits that need to be transported in the third phase. With the scheme of aggregate rate $K_1\frac{M^{b}}{\log M}$, we need $(CQ/K_1)M^{2-b}(!
 \log M)^2$ channel uses to complete the phase. Choosing  $M^{\frac{1}{2-b}}$, gives an aggregate throughput of $K_2n^{\frac{1}{2-b}}/(\log n)^2$ for the new scheme. This concludes the proof of Lemma~\ref{lem:lem2}.\hfill $\square$\\

We continue with the proofs of the lemmas introduced in the section:

\smallskip
\noindent\textit{Proof of Lemma~\ref{lem:sec4_1}:} The proof of the
lemma is a standard application of Chebyshev's inequality. Note that
the number of nodes in a given cell is a sum of $n$ i.i.d Bernoulli
random variables $B_i$, such that $\PP(B_i=1)=A_c$. Hence,
\begin{align*}
\PP\left(\sum_{i=1}^{n}B_i\geq (1+\delta)A_cn\right)&=\PP\left(e^{s\sum_{i=1}^{n}B_i}\geq e^{s(1+\delta)A_cn}\right)\\
&\leq \left(\EE[e^{sB_1}]\right)^n e^{-s(1+\delta)A_cn}\\
&=\left(e^{s}A_c+(1-A_c)\right)^n e^{-s(1+\delta)A_cn}\\
&\leq e^{-A_cn(s(1+\delta)-e^s+1)}\\
&= e^{-A_cn\Lambda_+(\delta)}
\end{align*}
where $\Lambda_+(\delta)=(1+\delta)\log(1+\delta)-\delta$ by
choosing $s=\log(1+\delta)$. The proof for the lower bound follows
similarly by considering the random variables $-B_i$. The conclusion
follows from the union bound. \hfill $\square$

\smallskip
\smallskip

\noindent\textit{Proof of Lemma~\ref{lem:sec4_2}:} Consider a node
$v$ in cluster $V$ operating under the 9-TDMA scheme in Figure 5.
The interfering signal received by this node from the simultaneously
operating clusters $\mathcal{U}_V$ is given by
$$
I_v=\sum_{U\in\,\mathcal{U}_V}\sum_{j\in U}H_{vj} \, X_j
$$
where $H_{vj}$ are the channel coefficients given by
(\ref{eq:ch_model}) and $X_j$ is the signal transmitted by node  $j$
which is located in a simultaneously operating cluster $U$. First
note that the signals $I_v$ and $I_{v^\prime}$ received by two
different nodes $v$ and $v^{\prime}$ in $V$ are uncorrelated since
the channel coefficients $H_{vj}$ and $H_{v^{\prime}j}$ are
independent for all $j$. The power of the interfering signal $I_v$
is given by
$$
P_I=\sum_{U\in\,\mathcal{U}_V}\sum_{j\in
U}\frac{GP_j}{(r_{vj})^{\alpha}}
$$
by using the fact that channel coefficients corresponding to
different nodes $j$ are independent. As illustrated by Figure 5, the
interfering clusters $\mathcal{U}_V$ can be grouped such that each
group $\mathcal{U}_V(i)$ contains $8i$ clusters or less and all
clusters in group $\mathcal{U}_V(i)$ are separated by a distance
larger than $(3i-1)\sqrt{A_c}$ from $V$ for $i=1,2,\dots$ where
$A_c$ is the cluster area. The number of such groups can be simply
bounded by the number of clusters $n/M$ in the network. Thus,
\begin{align}
P_I&<\sum_{i=1}^{n/M}\sum_{U\in\,\mathcal{U}_V(i)}\sum_{j\in\,U}\frac{GP_j}{((3i-1)\sqrt{A_c})^{\alpha}}\nonumber\\
&\leq\sum_{i=1}^{n/M}8i\frac{GP}{(3i-1)^{\alpha}}\label{eq:sum_int}
\end{align}
where we have used the fact that the powers of the signals are
bounded by $P_j \leq P \, A_c^{\alpha/2}/M,\,\forall j$. The sum in
(\ref{eq:sum_int}) is convergent for $\alpha>2$, thus is bounded by
a constant $K_{I_1}$. For $\alpha=2$, the sum can be bounded by
$K_{I_2}\log n$ where $K_{I_2}$ is a constant independent of $n$.
\begin{figure}
\begin{center}
\label{fig: int_9TDMA}
\input{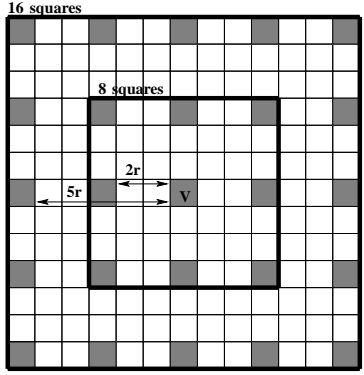}
\end{center}
\caption{Grouping of interfering clusters in the 9-TDMA Scheme.}
\end{figure}

\smallskip
\smallskip
\noindent\textit{Proof of Lemma~\ref{lem:sec4_5}:} We consider only
the case where the source cluster $S$ and the destination cluster
$D$ are not neighbors. The argument for the other case follows
similarly. The signal received by a destination node $d$ located in
cluster $D$ during MIMO transmission from source cluster $S$ is
given by
$$
Y_d = \sum_{s=1}^{M} H_{ds} \, X_s + Z_d
$$
where $X_s$ is the signal sent by a source node  $s \in S$
constrained to power $\frac{P(r_{SD})^{\alpha}}{M}$ and $Z_d$ is
$\sim\NC_\CC(0,N_0)$. The power of this signal is given by
\begin{align*}
\EE\left[|Y_d|^2\right]&=\sum_{s=1}^{M} |H_{ds}|^2
\,\frac{P(r_{SD})^{\alpha}}{M}  + N_0 \\
&=\sum_{s=1}^{M}\frac{GP}{M}
\,\left(\frac{r_{SD}}{r_{sd}}\right)^{\alpha}+N_0
\end{align*}
where we use the fact that all $H_{ds}$, $X_s$ and $Z_d$ are
independent. Observe that $r_{SD}-\sqrt{2A_c}\leq r_{sd}\leq
r_{SD}+\sqrt{2A_c}$, while $r_{SD}\geq 2\sqrt{A_c}$. These two
relations yield
$$
\left(\frac{\sqrt{2}}{\sqrt{2}+1}\right)^{\alpha}\leq\left(\frac{r_{SD}}{r_{sd}}\right)^{\alpha}\leq
\left(\frac{\sqrt{2}}{\sqrt{2}-1}\right)^{\alpha}
$$
which in turn yields the following lower and upper bounds for the
received power at each destination node
\begin{equation}
\label{eq: rec_pow}
P_1\equiv\left(\frac{\sqrt{2}}{\sqrt{2}+1}\right)^{\alpha}GP+N_0\leq\EE\left[|Y_d|^2\right]\leq
\left(\frac{\sqrt{2}}{\sqrt{2}-1}\right)^{\alpha}GP+N_0\equiv P_2.
\end{equation}\hfill $\square$

\section{Extended Networks}
\label{sec:extended}
\subsection{Bursty Hierarchical Scheme does better than Multihop for
$\alpha < 3$}

So far, we have considered {\em dense} networks, where the total
geographical area is fixed and the density of nodes increasing.
Another natural scaling is the {\em extended} case, where the
density of nodes is fixed and the area is increasing, a $\sqrt{n}
\times \sqrt{n}$ square. This models the situation where we want to
scale the network to cover an increasing geographical area.

As compared to dense networks, the distance between nodes is
increased by a factor of $\sqrt{n}$, and hence for the same transmit
powers, the received powers are all decreased  by a factor of
$n^{\alpha/2}$. Equivalently, by rescaling space, an extended
network can just be considered as a dense network on a unit area but
with the average power constraint per node reduced to
$P/n^{\alpha/2}$ instead of $P$.

Lemmas \ref{lem:lem1} and \ref{lem:lem2} state that the average
power per node required to run our hierarchical scheme in dense
networks is not the full power $P$ but $P/n$. In light of the
observation above, this immediately implies that when $\alpha =2$,
we can directly apply our scheme to extended networks and achieve a
{\em linear} scaling. For extended networks with $\alpha
> 2$, our scheme would not satisfy the equivalent power constraint
$P/n^{\alpha/2}$ and we are now in the power-limited regime (as
opposed to the degrees-of-freedom limited regime). However, we can
consider a simple ``bursty" modification of the hierarchical scheme
which runs the hierarchical scheme a fraction
$$\frac{1}{n^{\alpha/2-1}}$$
of the time with power $P/n$ per node and remains silent for the
rest of the time. This meets the given
average power constraint of $P/n^{\alpha/2}$, and achieves an
aggregate throughput of
$$\frac{1}{n^{\alpha/2-1}} \cdot n^{1-\epsilon} = n^{2-\alpha/2-\epsilon} \qquad \mbox{bits/second.}$$

Note that the quantity $n^{2-\alpha/2} = n^2 \cdot n^{-\alpha/2} $
can be interpreted as the total power transferred between a size $n$
transmit cluster  and a size $n$ receive cluster, $n^2$ node pairs
in all, with a power attenuation of $n^{-\alpha/2}$ for each node
pair. This power transfer is taking place at the top level of the
hierarchy (see Figure 3). The fact that the achievable rate is
proportional to the power transfer further emphasizes that our
scheme is power-limited rather than degrees-of-freedom limited in
extended networks.

Let us compare our scheme to multihop. For $\alpha <3$, it performs
strictly better than multihop, while for $\alpha > 3$, it performs
worse. Summarizing these observations, we have the following
achievability theorem for extended networks, the counterpart to
Theorem \ref{thm:thm1} for dense networks.

\begin{theorem}
\label{thm:extended} Consider an extended network on a $\sqrt{n}
\times \sqrt{n}$ square. There are two cases.
\begin{itemize}

\item {\bf $2 \le \alpha < 3$}: For every $\epsilon >0$, with high probability, an aggregate throughput:
$$ T(n)\geq Kn^{2-\alpha/2-\epsilon}$$
 is achievable in the network for all possible pairings between
sources and destinations. $K>0$ is a constant independent of $n$ and
the source-destination pairing.

\item {\bf $\alpha \ge 3$}: With high probability, an aggregate
throughput:
$$ T(n) \ge K\sqrt{n}$$
is achievable in the network for all possible pairings between
sources and destinations. $K>0$ is a constant independent of $n$ and
the source-destination pairing.
\end{itemize}

\end{theorem}
\smallskip
\smallskip

Note that because of the bursty transmission strategy, the
hierarchical scheme has a high peak-to-average power ratio. However,
although we talk in terms of time in the above discussion, such
burstiness can just as well be implemented over frequency with only
a fraction of the total bandwidth $W$ used. For example, this can be
implemented in an OFDM system, using a subset of the sub-carriers at
any one time, but putting more energy in the active sub-carriers.
This way, the peak power remains constant over time.

\subsection{Cutset Upper Bound for Random S-D Pairings}
Can we do better than the scaling in Theorem \ref{thm:extended}? So
far we have been considering arbitrary source-destination pairings
but clearly there are some pairings for which a much better scaling
can be achieved. For example, if the source-destinations are all
nearest neighbor to each other, then a linear capacity scaling can
be achieved for any $\alpha$. Thus, for the extended network case,
we need to narrow down the class of S-D pairings to prove a sensible
upper bound. In this section, we will therefore focus on {\em
random} S-D pairings, assuming that the pairs are chosen according
to a random permutation of the set of nodes, without any
consideration on node locations. We prove a high probability upper
bound that matches the achievability result in Theorem
\ref{thm:extended}, within a polynomial factor of arbitrarily small
exponent. Theorem \ref{thm:extended_ub} together with Theorem
\ref{thm:extended} identify therefore the capacity scaling law in
extended networks for {\em all} values of $\alpha \ge 2$. The rest
of the section is devoted to the proof of the theorem.

\begin{theorem}
\label{thm:extended_ub} Consider an extended network of $n$ nodes
with random source-destination pairing. For any $\epsilon>0$, the
aggregate throughput is bounded above by
$$ T(n) \leq \left \{ \begin{array}{ll} K^\prime\, n^{2-\alpha/2+\epsilon} &
2 \le \alpha  \leq 3 \\
K^\prime\, n^{1/2+\epsilon} & \alpha > 3 \end{array} \right.
$$
with high probability for a constant $K^\prime>0$ independent of
$n$.
\end{theorem}
\smallskip
\smallskip

Note that the hierarchical scheme is achieving near global
cooperation. In the context of dense networks, this yields a near
linear number of degrees of freedom for communication. In the
context of extended networks, in addition to the degrees of freedom
provided, this scheme allows almost all nodes in the network to
cooperate in transferring energy between any source-destination
pair. In fact, we saw that in extended networks with $\alpha > 2$,
our scheme is {\em power-limited} rather than degrees-of-freedom
limited. A natural place to look for a matching upper bound is to
consider a {\em cutset} bound on how much power can flow across the
network. Our proof of Theorem \ref{thm:extended_ub} is a careful
evaluation of such a cutset bound.


\smallskip
\smallskip

{\it Proof of Theorem \ref{thm:extended_ub}:} We start by
considering several properties that are satisfied with high
probability in the random network. The following lemma is similar in
spirit to Lemma~\ref{lem:sec4_1} for dense networks and can be
proved using a similar technique. In parallel to the dense case, it
forms the groundwork for our following discussion.

\begin{lemma}\label{lem:logn}
The random network with random source-destination pairing satisfies
the following properties with high probability:
\begin{itemize}
\item[a)] Let the network area be divided into $n$ squarelets of unit area. Then, there are less than $\log n$ nodes inside all squarelets.
\item[b)] Let the network area be divided into $\frac{n}{2\log n}$ squarelets each of area $2\log n$. Then, there is at least one node inside all squarelets.
\item[c)] Consider a cut dividing the network area into two equal halves. The number of communication requests with sources on the left-half network and destinations on the right-half network is between $\left((1-\delta)n/4,(1+\delta)n/4\right)$, for any $\delta>0$.
\end{itemize}
\end{lemma}

\begin{figure*}[tbp]
\begin{center}
\label{fig:cutset}
\input{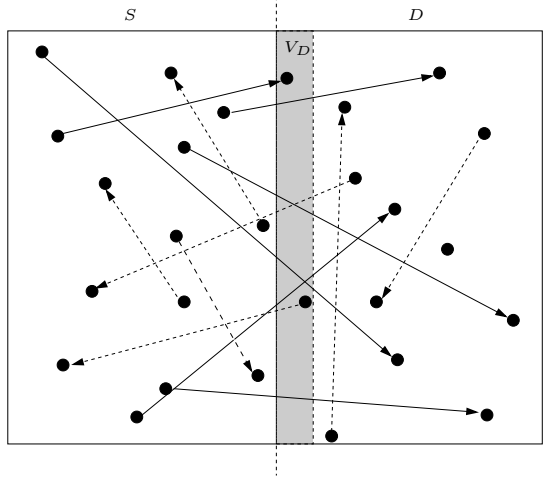}
\end{center}
\caption{The cut-set considered in the proof of Theorem
\ref{thm:extended_ub}. The communication requests that pass across
the cut from left to right are depicted in bold lines.}
\end{figure*}

We consider a cut dividing the $\sqrt{n}\times\sqrt{n}$ network area
into two equal halves (see Figure 6). We are interested in bounding
above the sum of the rates of communications passing through the cut
from left to right. By Part (c) of the lemma, this sum-rate is equal
to $1/4$'th of the total throughput $T(n)$ with high probability.
The maximum achievable sum-rate between these source-destination
pairs is bounded above by the capacity of the MIMO channel between
nodes $S$ located to the left of the cut and nodes $D$ located to
the right. Under the fast fading assumption, we have
\begin{equation} \label{cutset}
\sum_{k\in S, i\in D} R_{ik} \leq \max_{\substack{Q(H) \geq 0 \\
\EE(Q_{kk}(H)) \leq P, \, \forall k\in S}} \EE \left( \log \det(I +
H Q(H) H^*) \right),
\end{equation}
where
$$
H_{ik} = \frac{\sqrt{G} \; e^{j \, \theta_{ik}}}{r_{ik}^{\alpha/2}},
\quad k\in S, i\in D.
$$
$Q(\cdot)$ is a mapping from the set of possible channel
realizations $H$ to the set of positive semi-definite transmit
covariance matrices. The diagonal element $Q_{kk}(H)$ corresponds to
the power allocated to the $k$th node at channel state $H$. A
natural way to upper bound (\ref{cutset}) is by relaxing the
individual power constraint to a total transmit power constraint. In
the present context however, this is not convenient: some nodes in
$S$ are close to the cut and some are far apart, so the impact of
these nodes on the system performance is quite different. A total
transmit power constraint allows the transfer of power from the
nodes far apart to those nodes that are close to the cut, resulting
in a loose bound. Instead, we will relax the individual power
constraints to a total {\em weighted}  power constraint, where the
weight assigned to a node is set to be the total {\em received}
power on the other side of the cut per watt of transmit power from
that node. However, before doing that, we need to isolate the
contribution of some nodes in $D$ that are located very close to the
cut. Typically, there are few nodes on both sides of the cut that
are located at a distance as small as order $\frac{1}{\sqrt{n}}$
from the cut. If included, the contribution of these few pairs to
the total received power would be excessive, resulting in a loose
bound in the discussion below.

Let $V_D$ denote the set of nodes located on the $1\times\sqrt{n}$
rectangular area immediately to the right of the cut. Note that
there are no more than $\sqrt{n}\log n$ nodes in $V_{D}$ by Part (a)
of Lemma~\ref{lem:logn}. By generalized Hadamard's inequality, we
have
$$
\log \det\left(I + H Q(H) H^*\right)\leq \log\det\left(I + H^{(1)}
Q(H) H^{(1)*}\right)+\log\det\left(I + H^{(2)} Q(H) H^{(2)*}\right)
$$
where $H^{(1)}$ and $H^{(2)}$ are obtained by partitioning the
original matrix $H$: $H^{(1)}$ is the rectangular matrix with
entries $H_{ik}, k\in S, i\in V_D$ and $H^{(2)}$ is the rectangular
matrix with entries $H_{ik}, k\in S, i\in D\setminus V_D$. In turn,
(\ref{cutset}) is bounded above by
\begin{align} \nonumber
\sum_{k\in S, i\in D} R_{ik}\leq &
\max_{\substack{Q\left(H^{(1)}\right) \geq 0 \\
\EE\left(Q_{kk}\left(H^{(1)}\right)\right) \leq P, \, \forall k\in
S}} \EE \left( \log \det\left(I + H^{(1)} Q\left(H^{(1)}\right)
H^{(1)*}\right) \right)\\\label{cutset3}
&\qquad+\max_{\substack{Q\left(H^{(2)}\right) \geq 0 \\
\EE\left(Q_{kk}\left(H^{(2)}\right)\right) \leq P, \, \forall k\in
S}} \EE \left( \log \det\left(I + H^{(2)} Q\left(H^{(2)}\right)
H^{(2)*}\right) \right).
\end{align}
The first term in (\ref{cutset3}) can be easily upperbounded by
applying Hadamard's inequality once more or equivalently  by
considering the sum of the capacities of the individual MISO
channels between nodes in $S$ and each node in $V_D$. A discussion
similar to the proof of Theorem~\ref{thm:thm2} that makes use of the
fact that the minimum distance between any two nodes in the network
is larger than $\frac{1}{n^{1/2+\delta}}$ with high probability for
any $\delta>0$, yields the following upper bound for the first term
$$
\max_{\substack{Q\left(H^{(1)}\right) \geq 0 \\
\EE\left(Q_{kk}\left(H^{(1)}\right)\right) \leq P, \, \forall k\in
S}} \EE \left( \log \det\left(I + H^{(1)} Q\left(H^{(1)}\right)
H^{(1)*}\right) \right)\leq K^\prime\,\sqrt{n}(\log n)^2
$$
where $K^\prime>0$ is a constant independent of $n$.

The second term in (\ref{cutset3}) is the capacity of the MIMO
channel between nodes in $S$ and nodes in $D\setminus V_D$. This is
the term that dominates in (\ref{cutset3}) and thus its scaling
determines the scaling of (\ref{cutset}). The result is given by the
following lemma, which completes the proof of
Theorem~\ref{thm:extended_ub}.

\begin{lemma}
\label{lem:extended_ub} Let $P_{tot}(n)$ be the total power received
by all the nodes in $D\setminus V_D$, when nodes in $S$ are
transmitting {\em independent} signals at full power. Then for every
$\epsilon>0$, 
$$
\max_{\substack{Q\left(H^{(2)}\right) \geq 0 \\
\EE\left(Q_{kk}\left(H^{(2)}\right)\right) \leq P, \, \forall k\in
S}} \EE \left( \log \det\left(I + H^{(2)} Q\left(H^{(2)}\right)
H^{(2)*}\right) \right) \le n^\epsilon P_{tot}(n).
$$ 
Moreover, the scaling of the total received power can be evaluated
to be
$$
P_{tot}(n) \leq \left \{
\begin{array}{ll}
K^\prime\, n\, (\log n)^3 & \alpha =2 \\
K^\prime\, n^{2-\alpha/2}\, (\log n)^2 & 2 < \alpha <3 \\
K^\prime\,\sqrt{n} \, (\log n)^3 & \alpha=3 \\
K^\prime\,\sqrt{n} \, (\log n)^2& \alpha > 3 \end{array} \right.
$$
with high probability for a constant $K^\prime>0$ independent of
$n$.
\end{lemma}

\hfill $\square$\\

Lemma~\ref{lem:extended_ub} says two things of importance. First, it
says that independent signaling at the transmit nodes is sufficient
to achieve the cutset upper bound, as far as scaling is concerned.
There is therefore no need, in order for the transmit nodes to
cooperate, to do any sort of transmit beamforming. This is
fortuitous since our hierarchical MIMO performs only independent
signalling across the transmit nodes in the long-range MIMO phase.
Second, it identifies the total received power under independent
transmissions as the fundamental quantity limiting performance.
Depending on $\alpha$, there is a dichotomy on how this quantity
scales with the system size. This dichotomy can be interpreted as
follows.

The total received power is dominated either by the power
transferred between nodes near the cut (order $1$ distance) or by
the power transferred between nodes far away from the cut. There are
relatively fewer node {\em pairs} near the cut than away from the
cut (order $\sqrt{n}$ versus order $n^2$), but the channels between
the nodes near the cut are considerably stronger than between the
nodes far away from the cut. When the attenuation parameter $\alpha$
is less than $3$, the received power is dominated by transfer
between nodes far away from the cut. The hierarchical scheme, which
involves at the top level of the hierarchy MIMO transmissions
between clusters of size $n^{1-\epsilon}$ at distance $\sqrt{n}$
apart, achieves arbitrarily closely the required power transfer and
is therefore optimal in this regime. When $\alpha \ge 3$, the
received power in the cutset bound is dominated by the power
transfer by the nodes near the cut. This can be achieved by nearest
neighbor multihop and multihop is therefore optimal in this regime.

It should be noted that earlier works identified thresholds on
$\alpha$ above which nearest neighbor multihop is order-optimal ($\alpha>6$
first established in \cite{XK04}, then subsequently refined to hold for $\alpha > 5$
in \cite{JVK04}, $\alpha > 4.5$ in \cite{AJV06} and $\alpha > 4$ in \cite{XK06}
). All of them essentially use the same
cutset bound as we did. The fact that they did not identify the
tightest threshold (which we are showing to be $3$) is because their
upper bounds on the cutset bound are not tight.

{\it Proof of Lemma~\ref{lem:extended_ub}:} We are interested in the
scaling of the MIMO capacity,
\begin{equation}\label{MIMO_cap}
\max_{\substack{Q\left(H^{(2)}\right) \geq 0 \\
\EE\left(Q_{kk}\left(H^{(2)}\right)\right) \leq P, \, \forall k\in
S}} \EE \left( \log \det\left(I + H^{(2)} Q\left(H^{(2)}\right)
H^{(2)*}\right) \right).
\end{equation}
Let us rescale each column $k$ of the matrix by the (square root of
the) {\it total received power} on the right from source node $k$ on
the left. Let indeed $P_k$ denote the total received power in
$D\setminus V_D$ of the signal sent by user $k \in S$:
$$
P_k = P \, G \, \sum_{i \in D\setminus V_D} r_{ik}^{-\alpha} := P \,
G \, d_k.
$$
The expression (\ref{MIMO_cap}) is then equal to
$$
\max_{\substack{\tilde{Q}\left(\tilde{H}\right) \geq 0 \\
\EE\left(\tilde{Q}_{kk}\left(\tilde{H}\right)\right) \leq P_k, \,
\forall k\in S}} \EE \left(  \log \det \left( I + \tilde{H}
\tilde{Q}\left(\tilde{H}\right) \tilde{H}^*\right) \right),
$$
where
$$
\tilde{H}_{ik} = \frac{e^{j \, \theta_{ik}}} {r_{ik}^{\alpha/2}} \,
\frac{1}{\sqrt{d_k}},\qquad i\in D\setminus V_D, k\in S.
$$
The above expression is in turn bounded above by
$$
\max_{\substack{\tilde{Q}(\tilde{H}) \geq 0 \\ \EE(\Tr\tilde{Q}(H))
\leq P_{tot}(n)}} \EE \left(  \log \det ( I + \tilde{H}
\tilde{Q}(\tilde{H}) \tilde{H}^*) \right),
$$
where $P_{tot}(n) = \sum_{k \in S} P_k = P \, G \sum_{k \in S,\, i
\in D\setminus V_D} r_{ik}^{-\alpha}$.

Let us now define, for given $n \geq 1$ and $\varepsilon>0$, the set
$$
B_{n,\varepsilon} = \{ \Vert \tilde{H} \Vert^2 > n^\varepsilon \},
$$
where $\Vert A \Vert$ denotes the largest singular value of the
matrix $A$. Note that the matrix $\tilde{H}$ is better conditioned
than the original channel matrix $H^{(2)}$: all the diagonal
elements of $\tilde{H}\tilde{H}^*$ are roughly of the same order (up
to a factor $\log n$), and it can be shown that there exists
$K_1^{\prime}>0$ such that
$$
\EE( \Vert \tilde{H} \Vert^2) \leq K_1^{\prime} \, (\log n)^3
$$
for all $n$. In Appendix \ref{app:largest_ev}, we show the following
more precise  statement.

\begin{lemma} \label{lem:largest_ev}
For any $\varepsilon>0$ and $p \geq 1$, there exists
$K_1^{\prime}>0$ such that for all $n$,
$$
\PP(B_{n,\varepsilon}) \leq \frac{K_1^{\prime}}{n^p}.
$$
\end{lemma}
\smallskip
\smallskip

It follows that
\begin{align}
\max_{\substack{\tilde{Q}(\tilde{H}) \geq 0 \\
\EE(\Tr\tilde{Q}(\tilde{H})) \leq P_{tot}(n)}} &\EE \left(
\log \det ( I + \tilde{H} \tilde{Q}(\tilde{H}) \tilde{H}^*) \right) \nonumber\\
& \leq  \max_{\substack{\tilde{Q}(\tilde{H}) \geq 0 \\
\EE(\Tr\tilde{Q}(\tilde{H})) \leq P_{tot}(n)}} \EE \left( \log \det
( I + \tilde{H} \tilde{Q}(\tilde{H}) \tilde{H}^*) \,
1_{B_{n,\varepsilon}} \right)\nonumber\\ &~~~~+
\max_{\substack{\tilde{Q}(\tilde{H}) \geq 0 \\
\EE(\Tr\tilde{Q}(\tilde{H})) \leq P_{tot}(n)}}\EE \left( \Tr
(\tilde{H} \tilde{Q}(\tilde{H}) \tilde{H}^*) \,
1_{B_{n,\varepsilon}^c} \right) \label{cutset2}
\end{align}
The first term in (\ref{cutset2}) refers to the event that the
channel matrix $\tilde{H}$ is accidentally ill-conditioned. Since
the probability of such an event is polynomially small by Lemma
\ref{lem:largest_ev}, the contribution of this first term is
actually negligible. In the second term in (\ref{cutset2}), the
matrix $\tilde{H}$ is well conditioned, and this term is actually
proportional to the maximum power transfer from left to right.
Details follow below.

For the first term in (\ref{cutset2}), we use Hadamard's inequality
and obtain
\begin{align*}
\max_{\substack{\tilde{Q}(\tilde{H}) \geq 0 \\
\EE(\Tr\tilde{Q}(\tilde{H})) \leq P_{tot}(n)}}&\EE \left(
\log \det ( I + \tilde{H} \tilde{Q}(\tilde{H}) \tilde{H}^*) \, 1_{B_{n,\varepsilon}} \right)\\
& \leq \max_{\substack{\tilde{Q}(\tilde{H}) \geq 0 \\
\EE(\Tr\tilde{Q}(\tilde{H})) \leq P_{tot}(n)}} \EE \left( \sum_{i
\in D\setminus V_D} \log ( 1 + \tilde{H}_i \tilde{Q}(\tilde{H})
\tilde{H}_i^*) \, \Bigg| \, B_{n,\varepsilon} \right)
\PP(B_{n,\varepsilon})
\end{align*}
where $\tilde{H}_i$ is the $i^{th}$ row of $\tilde{H}$. By Jensen's
inequality, this expression in turn is bounded above by
\begin{align*}
& \max_{\substack{\tilde{Q}(\tilde{H}) \geq 0 \\
\EE(\Tr\tilde{Q}(\tilde{H})) \leq P_{tot}(n)}}   \sum_{i \in D\setminus V_D} \log \left( 1 +  \EE\left(\Vert \tilde{H}_i \Vert^2 \, \Tr \tilde{Q}(\tilde{H})\Big| \, B_{n,\varepsilon}\right) \right) \PP(B_{n,\varepsilon})\\
& \leq \max_{\substack{\tilde{Q}(\tilde{H}) \geq 0 \\
\EE(\Tr\tilde{Q}(\tilde{H})) \leq P_{tot}(n)}}  \sum_{i \in D\setminus V_D} \log \left( 1 +  \EE\left(\Vert \tilde{H}_i \Vert^2 \, \Tr \tilde{Q}(\tilde{H})\right)\Big{/}\PP(B_{n,\varepsilon}) \right) \PP(B_{n,\varepsilon})\\
& \leq  n \log \left(1 + \frac{n \,
P_{tot}(n)}{\PP(B_{n,\varepsilon})} \right) \,
\PP(B_{n,\varepsilon}),
\end{align*}
since
$$
\Vert \tilde{H}_i \Vert^2 = \sum_{k \in S} r_{ik}^{-\alpha} \,
\frac{1}{d_k} \leq \sum_{k \in S} 1 \leq n.
$$
The fact that the minimum distance between the nodes in $S$ and
$D\setminus V_D$ is at least $1$ yields $P_{tot}(n)\leq PG n^{2}$.
Noting that $x \mapsto x \log(1+ 1/x)$ is increasing on $[0,1]$ and
using Lemma \ref{lem:largest_ev}, we obtain finally that for any $p
\geq 1$, there exists $K_1^\prime>0$ such that
$$
\max_{\substack{\tilde{Q}(\tilde{H}) \geq 0 \\
\EE(\Tr\tilde{Q}(\tilde{H})) \leq P_{tot}(n)}} \EE \left( \log \det
( I + \tilde{H} \tilde{Q}(\tilde{H}) \tilde{H}^*) \,
1_{B_{n,\varepsilon}} \right) \leq K_1^\prime \, n^{1-p} \log \left(
1 + \, \frac{n^{3+p}}{K_1^\prime} \right),
$$
which decays polynomially to zero with arbitrary exponent as $n$
tends to infinity.

For the second term in (\ref{cutset2}), we simply have
\begin{align*}
\max_{\substack{\tilde{Q}(\tilde{H}) \geq 0 \\
\EE(\Tr\tilde{Q}(\tilde{H})) \leq P_{tot}(n)}}\EE \left( \Tr
(\tilde{H} \tilde{Q}(\tilde{H}) \tilde{H}^*) \,
1_{B_{n,\varepsilon}^c} \right) &\leq
\max_{\substack{\tilde{Q}(\tilde{H}) \geq 0 \\
\EE(\Tr\tilde{Q}(\tilde{H})) \leq P_{tot}(n)}}\EE \left( \Vert
\tilde{H} \Vert^2
\Tr\tilde{Q}(\tilde{H}) \, 1_{B_{n,\varepsilon}^c} \right)\\
&\leq n^\varepsilon \, P_{tot}(n).
\end{align*}
The last thing that needs therefore to be checked is the scaling of $P_{tot}(n)$ stated in Lemma~\ref{lem:extended_ub}.\\

\begin{figure*}[tbp]
\begin{center}
\label{fig: binning}
\input{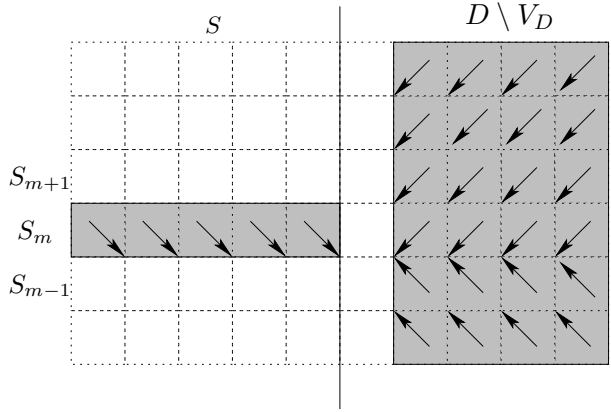}
\end{center}
\caption{The displacement of the nodes inside the squarelets to
squarelet vertices, indicated by arrows.}
\end{figure*}

Let us divide the network area into $n$ squarelets of area $1$. By
Part (a) of Lemma~\ref{lem:logn}, there are no more than $\log n$
nodes in each squarelet with high probability. Let us consider
grouping the squarelets of $S$ into $\sqrt{n}$ rectangular areas
$S_m$ of height $1$ and width $\sqrt{n}$ as shown in Figure 8. Thus,
$S=\bigcup_{m=1}^{\sqrt{n}}S_m$. We are interested in bounding above
$$
P_{tot}(n)=P G\sum_{k\in S}d_k=P G\sum_{m=1}^{\sqrt{n}}\sum_{k\in
S_m}d_k.
$$
Let us consider
\begin{equation}\label{S_m}
\sum_{k\in S_m}d_k=\sum_{k\in S_m,i \in D\setminus V_D}
r_{ik}^{-\alpha}
\end{equation}
for a given $m$. Note that if we move the points that lie in each
squarelet of $S_m$ together with the nodes in the squarelets of
$D\setminus V_D$ onto the squarelet vertex as indicated by the
arrows in Figure 7, all the (positive) terms in the summation in
(\ref{S_m}) can only increase since the displacement can only
decrease the Euclidean distance between the nodes involved. Note
that the modification results in a regular network with at most
$\log n$ nodes at each squarelet vertex on the left and at most
$2\log n$ nodes at each squarelet vertex on the right. Considering
the same reasoning for all rectangular slabs
$S_m,\,,m=1,\dots,\sqrt{n}$ allows to conclude that $P_{tot}(n)$ for
the random network is with high probability less than the same
quantity computed for a regular network with $\log n$ nodes at each
left-hand side vertex and $2\log n$ nodes at each right-hand side
vertex.

The most convenient way to index the node positions in the resulting
regular network is to use double indices. The left-hand side nodes
are located at positions $(-k_x+1,k_y)$ and those on the right at
positions $(i_x,i_y)$ where $k_x,k_y,i_x,i_y=1,\ldots,\sqrt{n}$, so
that
$$
\tilde{H}_{ik} = \frac{e^{j \,
\theta_{ik}}}{((i_x+k_x-1)^2+(i_y-k_y)^2)^{\alpha/4}} \,
\frac{1}{\sqrt{d_{k_x,k_y}}}
$$
and
\begin{equation}\label{dk_reg}
d_{k_x,k_y} = \sum_{i_x,i_y=1}^{\sqrt{n}}
\frac{1}{((i_x+k_x-1)^2+(i_y-k_y)^2)^{\alpha/2}}
\end{equation}
which yields the following upper bound for $P_{tot}(n)$ of the
random network,
\begin{align}
P_{tot}(n)&\leq  2(\log n)^2 P
G\sum_{k_x,k_y=1}^{\sqrt{n}}d_{k_x,k_y}.\label{ptot_ub}
\end{align}

The following lemma establishes the scaling of $d_{k_x,k_y}$ defined
in (\ref{dk_reg}).
\begin{lemma}
\label{lem:dk_scaling} There exist constants $K_2^\prime,
K_3^\prime>0$ independent of $k_x$, $k_y$ and $n$ such that
$$
d_{k_x,k_y}\leq\left\{
\begin{array}{ll}
K_2^\prime\,\log n & \text{if } \alpha=2,\\
K_2^\prime\,k_x^{2-\alpha} & \text{if } \alpha>2,
\end{array} \right.
$$
and
$$
d_{k_x,k_y}\geq
K_3^\prime\,k_x^{2-\alpha}\qquad\textrm{for}\qquad\alpha\geq2.
$$
\end{lemma}

The rigorous proof of the lemma is given at the end of
Appendix~\ref{app:largest_ev}. A heuristic way of thinking about the
approximation
$$ d_{k_x,k_y} \approx k_x^{2-\alpha}$$
can be obtained through Laplace's principle. The summation in
$d_{k_x,k_y}$ scales the same as the maximum term in the sum times
the number of terms which have roughly this maximum value. The
maximum term is of the order of $1/k_x^{\alpha}$. The terms that
take on roughly this value are those for which $i_x$ runs from $1$
to the order of $k_x$ and $i_y$ runs from $k_y$ to $k_y$ plus or
minus the order of $k_x$. There are roughly $k_x^2$ such terms.
Hence $d_{k_x,k_y} \approx 1/k_x^\alpha \cdot k_x^2 =
k_x^{2-\alpha}$.

We can now use the upper bound given in the above lemma to yield:
$$
\sum_{k_x,k_y=1}^{\sqrt{n}}d_{k_x,k_y}\leq\left\{
\begin{array}{ll}
K_4^\prime\,n\log n & \text{if } \alpha=2,\\
K_4^\prime\,n^{2-\alpha/2} & \text{if } 2<\alpha\leq 3,\\
K_4^\prime\,\sqrt{n}\log n & \text{if } \alpha=3,\\
K_4^\prime\,\sqrt{n} & \text{if } \alpha>3
\end{array} \right.
$$
for another constant $K_4^\prime>0$ independent of $n$. This upper
bound combined with (\ref{ptot_ub}) completes the proof of
Lemma~\ref{lem:extended_ub}.\hfill $\square$

\section{Discussions on the Model and the Results}
\label{sec:discussions}

In this section, we point out the scope and limitations of the model
and the results.

\subsection{Scaling Laws vs Performance Analysis}

We should emphasize that the focus in this paper, as well as in
\cite{GK00} and the follow-up works, is on scaling laws, i.e.
scaling of the aggregate throughput in the limit when the number of
users gets large. The main advantage of studying scaling laws is to
highlight qualitative and architectural properties of the system
without getting bogged down by too many details. For example, the
linear scaling law in dense networks we derived highlights the fact
that interference limitation in the Gupta-Kumar scaling is not a
fundamental one and can be alleviated by more complicated physical
layer processing.

It is important to distinguish between such scaling law study and
the design and performance analysis of a scheme for a network with a
given number of users. While scaling law results provide some
architectural guidelines on how to design schemes that scale well,
detailed design and performance analysis would involve tuning of
many parameters and improvements of the scheme to optimize the
pre-constant in the system throughput.  For example, our scheme
quantizes the received analog signal at each node and forward the
bits to the final destination, but the quantized bits are correlated
across the receive nodes and hence a reduction in the overhead can
be achieved by doing some Slepian-Wolf coding. Such work is beyond
the scope of the present paper.

We studied two different scaling laws in this paper, one for dense
and one for extended networks. Given a network with a specific
number of nodes occupying a specific area, a natural question is: is
this network best described by the dense scaling regime or the
extended scaling regime? What our results say is that a better
delineation is in terms of whether we are in the degree-of-freedom
limited or power(coverage)-limited regime, because this is what will have
architectural implications for the communication scheme (for
example, whether bursty transmissions are required). To get a sense
of the operating regime a given network is in, our results
suggest a rule-of-thumb quantity that can be calculated: the total
received SNR per node, total over all the transmit powers of the
nodes in the network. If this quantity is much larger than $0$ dB,
then the network is in the degree-of-freedom limited regime;
otherwise it is in the power-limited regime.

\subsection{Far-field Assumption in Dense Scaling}
One potential concern with the dense scaling is that the far-field
assumption will eventually break down as the number of nodes gets
too large. In practice, the typical separation between nodes is so
much larger than the carrier wavelength that the number of nodes for
which the far-field assumption fails is humongous, i.e. there is a
clear separation between the large and the small spatial scales.
Consider the following numerical example: suppose the area of
interest is $1$ square km, well within the communication range of
many radio devices. With a carrier frequency of $3$ GHz, the carrier
wavelength is $0.1$m. Even with a very large system size of
$n=10000$ nodes, the typical separation between nearest neighbors is
$10$ m, very much in the far-field. Under free-space propagation and
assuming unit transmit and receive antenna gains, the attenuation
given by Friis' formula (\ref{eq:friis}) is about $10^{-6}$, much
smaller than unity. At the same time, the total received SNR per
node (assuming transmit power $P$  of $1$ mW per node, thermal noise
$N_0$ at $-174$ dBm, a bandwidth $W$ of $10$ MHz and noise figure
NF= $10$ dB) is $84$ dB, very much in the degree-of-freedom limited
regime\footnote{$\SNR_{\rm dB} = P_{\rm dBm} + 10 \, \log_{10} n +
{\rm pathloss}_{\rm dB} - (N_0)_{\rm dBm} - 10 \log_{10} W -{\rm
NF}_{\rm dB}.$}. (Looking at even only one point-to-point link at
distance $1$ km, the received SNR is $34$ dB). Hence, this example
gives evidence that there are networks for which simultaneously the
number of nodes is large, the far-field assumption holds and the
received SNR across the network is high. However, a careful performance
analysis of the pre-constants is required to confirm that linear
scaling of our scheme has already kicked in and our scheme indeed
outperforms multi-hop in this parameter range.
Nevertheless, we do believe that the
linear scaling obtained here also applies for a relatively small
number of nodes. The intuition for this is that our strategy relies
on the use of MIMO communication, whose linear capacity scaling has
never been disputed in the range of a small number of antennas.

\subsection{$d$-dimensional networks}

We have focused on the 2D setting, where the nodes are on the
plane, but our results generalize naturally to networks where nodes
live in $d$-dimensional space. For dense networks, linear scaling is
achievable whenever $\alpha > d$, i.e.~whenever spatial reuse is
possible. For extended networks, the scaling exponent $e^*_d(\alpha)$ is given by:
$$
e_d^*(\alpha) = \left \{ \begin{array}{cc}
2- \frac{\alpha}{d} & d \le \alpha \le d+1\\ 1-\frac{1}{d} & \alpha > d+1 \end{array} \right.
$$
For $\alpha$ between $d$ and $d+1$, hierarchical MIMO achieves the
optimal scaling, and for $\alpha > d+1$, nearest neighbor multihop
is optimal.

\subsection{Transport Capacity}

Let us finally mention that a more general measure of network performance
has been introduced in \cite{GK00}: the {\em transport capacity} of a network,
defined as the maximum number of bits exchanged in the network per second,
weighted by their travelled distances. From an upper bound on transport
capacity, one can easily deduce an upper bound on the aggregate throughput
for the special case where the source-destination pairs are chosen at random and
communicating at a common rate, which is the traffic requirement considered
in the current paper. But the interest in an {\em upper bound} on transport
capacity lies in the fact that it applies to more general communication
scenarios. Reciprocally, it has been shown recently in \cite{OLP07} that for
a network with a random placement of nodes, there is a natural way to deduce
an upper bound on transport capacity from an upper bound on throughput, by
studying cutset bounds over multiple cuts (as first suggested in \cite{XK04}).
Applying this technique to the present result leads to the following
conclusion: the transport capacity $T_c(n)$ of the extended network is upper bounded by
\begin{itemize}
\item  $T_c(n)\leq\,K^\prime\, n^{2.5-\alpha/2+\varepsilon}$, for $2 \leq \alpha \leq 3$,
\item  $T_c(n)\leq\,K^\prime\, n^{1+\varepsilon}$, for $\alpha > 3$,
\end{itemize}
for any $\varepsilon>0$, where $K^\prime>0$ is a constant independent of $n$. Note that these
scaling laws for the transport capacity are also achievable within a factor of $n^\varepsilon$.

\section{Conclusions}
\label{sec:conclusions}

In point-to-point communication, performance is limited by either
the power or the degrees of freedom (bandwidth and
number of antennas) available, depending on whether the link is operating at
low or high signal-to-noise ratio. In a network with multiple
source-destination pairs, performance can further be limited by the
interference between simultaneous transmission of information. In
this paper, we have shown that by achieving near global MIMO
cooperation between nodes without introducing significant cooperation overhead,
interference can be successfully removed as a limitation, at least
as far as scaling laws are concerned. Moreover, such near-global
MIMO cooperation also allows the maximum transfer of energy between
all source-destination pairs, provided that the path loss across the
network is not too much. This implies that in degrees-of-freedom
limited scenarios, such as in dense networks or extended networks
with path loss exponent $\alpha =2$, the full degrees of freedom in
the network can be shared among all nodes and a linear capacity
scaling can be achieved. In power-limited scenarios but with low
attenuation, such as extended networks with $\alpha$ between $2$ and
$3$, our scheme achieves the optimal (power-limited) capacity
scaling law.

The key ideas behind our scheme are:

\begin{itemize}

\item using MIMO for long-range communication to achieve spatial
multiplexing;
\item local transmit and receive cooperation to maximize spatial reuse;
\item setting up the intra-cluster cooperation such that it is yet
another digital communication problem, but in a smaller network,
thus enabling a hierarchical cooperation architecture.

\end{itemize}


%
%



\section*{Acknowledgment}
David Reed raised the question of whether a linear capacity scaling
is possible, that provided part of the impetus for this research.
The authors would also like to thank to Emre Telatar, Shuchin Aeron
and Venkatesh Saligrama for many helpful discussions. The work of
Ayfer {\"O}zg{\"u}r was supported by Swiss NSF grant Nr
200021-10808. Part of the work of Olivier L{\'e}v{\^e}que was
performed when he was with the Electrical Engineering Department at
Stanford University, supported by Swiss NSF grant Nr PA002-108976.
The work of David Tse was supported by the U.S. National Science
Foundation via an ITR grant: "The 3R's of Spectrum Management:
Reuse, Reduce and Recycle".

\appendices

\section{Linear scaling law for the MIMO gain under fast fading assumption}
\label{app:MIMO} \noindent\textit{Proof of Lemma~\ref{lem:sec4_3}:}
The $M \times M$ MIMO channel between two clusters $S$ and $D$ is
given by $Y=H X + Z$, where $H_{ik}$ are given in
(\ref{eq:ch_model}). Recall that $Z=(Z_k)$ is uncorrelated
background noise plus interference  at the receiver nodes. Assume
that the transmitted signals $X=(X_i)$ are from an i.i.d.
$\sim\NC_\CC(0,\sigma^2)$ randomly chosen codebook with
$$
\sigma^2=\frac{P (r_{SD})^\alpha}{M}.
$$
It is well known that the achievable mutual information is lower
bounded by assuming that the interference-plus-noise $Z$ is i.i.d.
Gaussian. (see for example Theorem 5 of \cite{ET06} for a precise
statement and proof of this in the MIMO case.) With our transmission
strategy in the MIMO phase, there exists $b>a>0$ with $a$ and $b$
independent of $n$, such that $r_{ik}^{-\alpha/2}=r_{SD}^{-\alpha/2}
\, \rho_{ik}$, where all $\rho_{ik}$ lie in the interval $[a,b]$
both in the cases when $S$ and $D$ are neighboring clusters or not.

By assuming perfect channel state information at the receiver side,
the mutual information of the above MIMO channel is given by
\begin{equation}\label{mutinf}
I(X; Y, H) \ge \EE \left(\log \det \left( I + \frac{\sigma^2}{N} \,
HH^* \right) \right) = \EE \left( \log \det \left(I + \frac{\SNR}{M}
\, F F^* \right) \right),
\end{equation}
where $\SNR =\frac{ G P}{N}$ ($N$= total interference-plus-noise
power) and $F_{ik} = \rho_{ik} \, \exp(j \, \theta_{ik})$. Let
$\lambda$ be chosen uniformly among the $M$ eigenvalues of
$\frac{1}{M} FF^*$. The above mutual information may be rewritten as
$$
I(X; Y, H) \ge M \, \EE ( \log (1+ \SNR \,\lambda)) \geq M \, \log
(1+ \SNR \,t) \, \PP(\lambda>t),
$$
for any $t \geq 0$. By the Paley-Zygmund inequality, if $0 \leq t <
\EE(\lambda)$, we have
$$
\PP(\lambda>t) \geq \frac{(\EE(\lambda)-t)^2}{\EE(\lambda^2)}.
$$
We therefore need to compute both $\EE(\lambda)$ and
$\EE(\lambda^2)$. We have,
\begin{align*}
\EE(\lambda) &= \frac{1}{M} \, \EE\left( \Tr\left(\frac{1}{M} \, FF^*\right)\right)\\
& = \frac{1}{M^2} \sum_{i,k=1}^M \EE(|F_{ik}|^2) \\
&= \frac{1}{M^2} \sum_{i,k=1}^M \rho_{ik}^2 \geq a^2
\end{align*}
and
\begin{eqnarray*}
\EE(\lambda^2) & = & \frac{1}{M} \EE \left(\Tr \left(\frac{1}{M^2}
FF^* FF^* \right)
\right)\\
&=& \frac{1}{M^3} \sum_{iklm=1}^M \EE (F_{ik} \overline{F_{lk}} F_{lm} \overline{F_{im}})\\
& \leq & \frac{2}{M^3} \sum_{ikm=1}^M \EE_r(|F_{ik}|^2) \, \EE_r(|f_{im}|^2)\\
&=& \frac{2}{M^3} \sum_{ikm=1}^M \rho_{ik}^2 \, \rho_{im}^2 \leq 2
b^4,
\end{eqnarray*}
so $\EE(\lambda) \geq a^2$ and $\EE(\lambda^2) \leq 2 b^4$. This
leads us to the conclusion that for any $t<a$, we have
\begin{equation}\label{MIMO_gain}
I(X;Y,H) \geq M \, \log (1+\SNR\,t) \, \frac{(a^2-t)^2}{2b^4},
\end{equation}
Choosing e.g.~$t=a/2$ shows that $I(X;Y,H)$ grows at least linearly
with $M$.\hfill $\square$

\smallskip
\smallskip
\textit{Lemma I.1 (Paley-Zygmund Inequality)} Let $X$ be a
non-negative random variable such that $\EE(X^2)<\infty$. Then for
any $t \geq 0$  such that $t < \EE(X)$, we have
$$
\PP(X > t) \geq \frac{(\EE(X)-t)^2}{\EE(X^2)}.
$$

\smallskip
\smallskip
\textit{Proof:} By the Cauchy-Schwarz inequality, we have for any $t
\geq 0$:
$$
\EE( X \, 1_{X>t} ) \leq \sqrt{\EE(X^2) \, \PP(X>t)}.
$$
and also, if $t < \EE(X)$,
$$
\EE( X \, 1_{X>t} ) = \EE(X) - \EE( X \, 1_{X \leq t}) \geq \EE(X) -
t >0.
$$
Therefore,
$$
\PP(X>t) \geq \frac{(\EE(X)-t)^2}{\EE(X^2)}.
$$
\hfill $\square $

Note that the achievability results in this paper can be extended to
the slow fading case, provided that Lemma~\ref{lem:sec4_3} can be
proved in the slow fading setting. In that case, one would need to
show that the expression inside the expectation in (\ref{mutinf})
concentrates around its mean exponentially fast in $M$. However,
another difficulty might arise from the lack of averaging of the
phases in the interference term, which leads to a non-spatially
decorrelated noise term $Z$. Although proving the result might
require some technical effort, we believe it holds true, due to the
self-averaging effect of a large number of independent random
variables.

\section{Achievable Rates on Quantized Channels}
\label{app:quant_MIMO} In order to conclude the discussion on the
throughput achieved by our scheme, we need to show that the
quantized MIMO channel achieves the same spatial multiplexing gain
as the MIMO channel. In Theorem~II.1, we give a simple
achievability region for general quantized channels. Note that a stronger
result is established in \cite[Theorem~3]{KGG05} that implies Theorem~II.1 as a special case. The required result for
the quantized MIMO channel is then found as an easy application of
Theorem~II.1. We start by formally defining the general quantized
channel problem in a form that is of interest to us and proceed with
several definitions that will be needed in the sequel.

Let us consider a discrete-time memoryless channel with single input
of alphabet $\mathcal{X}$ and $M$ outputs of respective alphabets
$\mathcal{Y}_1,\dots,\mathcal{Y}_M$. The channel is statistically
described by a conditional probability distribution
$p(y_1,\dots,y_M|x)$ for each
$y_1\in\mathcal{Y}_1,\dots,y_M\in\mathcal{Y}_M$ and
$x\in\mathcal{X}$. The outputs of the channel are to be followed by
quantizers which independently map the output alphabets
$\mathcal{Y}_1,\dots,\mathcal{Y}_M$ to the respective reproduction
alphabets $\hat{\mathcal{Y}}_1,\dots,\hat{\mathcal{Y}}_M$. The aim
is to recover the transmitted information through the channel by
observing the outputs of the quantizers. Communication over the
channel takes place in the following manner: A message $W$, drawn
from the index set $\{1,2,\dots, L\}$ is encoded into a codeword
$X^m(W)\in\mathcal{X}^m$, which is received as $M$ random sequences
$(Y_1^m,\dots,Y_M^m)\sim p(y_1^m,\dots,y_M^m|x^m)$ at the outputs of
the channel. The quantizers themselves consist of encoders and
decoders, where the $j$'th encoder describes its corresponding
received sequence $Y_j^m$ by an index
$I_j(Y_j^m)\in\{1,2,\dots,L_j\}$, and decoder $j$ represents $Y_j^m$
by an estimate $\hat{Y}_j^m(I_j)\in\hat{\mathcal{Y}}_j^m$. The
channel decoder then observes the reconstructed sequences
$\hat{Y}_1^m,\dots,\hat{Y}_M^m$ and guesses the index $W$ by an
appropriate decoding rule
$\hat{W}=g(\hat{Y}_1^m,\dots,\hat{Y}_M^m)$. An error occurs if
$\hat{W}$ is not the same as the index $W$ that was transmitted. The
complete model under investigation is shown in Fig.~7. An $(L;
L_1,\dots,L_M; m)$ code for this channel is a joint $(L,m)$ channel
and $M$ quantization codes $(L_1,m),\dots,(L_M,m)$; more
specifically, is two sets of encoding and decoding functions, the
first set being the channel encoding function
$X^m:\{1,2,\dots,L\}\rightarrow\mathcal{X}^m$ and the channel
decoding function
$g:\hat{\mathcal{Y}}_1^m\times\dots\times\hat{\mathcal{Y}}_M^m\rightarrow\{1,2,\dots,L\}$,
and the second set consists of the encoding functions
$I_j:\mathcal{Y}_j^m\rightarrow\{1,2,\dots,L_j\}$ and decoding
functions
$\hat{Y}_j^m:\{1,2,\dots,L_j\}\rightarrow\hat{\mathcal{Y}}_j^m$ for
$j=1,\dots,M$ used for the quantizations. We define the (average)
probability of error for the $(L;L_1,\dots,L_M; m)$ code by
$$
P_e^m=\frac{1}{L}\sum_{k=1}^{L}\mathbb{P}(\hat{W}\neq k~|~W=k).
$$
A set of rates $(R; R_1,\dots,R_M)$ is said to be achievable if
there exists a sequence of \newline
$(2^{mR};2^{mR_1},\dots,2^{mR_M};m)$ codes with $P_e^m\rightarrow 0$
as $m\rightarrow\infty$. Note that determining achievable rates $(R;
R_1,\dots,R_M)$ is not a trivial problem, since there is trade-off
between maximizing $R$ and minimizing $R_1,\dots,R_M$.

\begin{figure*}[tbp]
\begin{center}
\label{fig: quant_ch}
\input{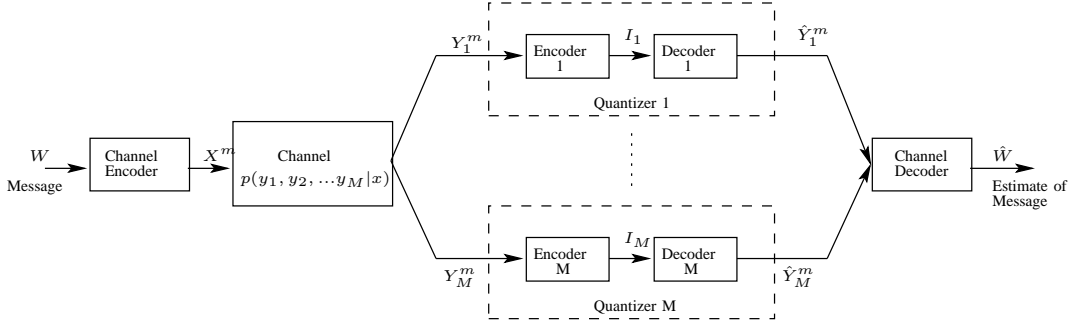}
\end{center}
\caption{The Quantized Channel Problem.}
\end{figure*}

\smallskip
\smallskip
\textit{Theorem~II.1 (Achievability for the Quantized Channel
Problem)} Given a probability distribution $q(x)$ on $\mathcal{X}$
and $M$ conditional probability distributions $q_j(\hat{y}_j|y_j)$
where $y_j\in\mathcal{Y}_j$ and $\hat{y}_j\in\hat{\mathcal{Y}}_j$
and $j=1,\dots,M$; all rates $(R;R_1,\dots,R_M)$ such that
$R<I(X;\hat{Y}_1,\dots,\hat{Y}_M)$ and $R_j>I(Y_j;\hat{Y}_j)$ are
achievable. Specifically, given any $\delta>0$, $q(x)$ and
$q_j(\hat{y}_j|y_j)$, together with rates
$R<I(X;\hat{Y}_1,\dots,\hat{Y}_M)$ and $R_j>I(Y_j;\hat{Y}_j)$ for
$j=1,\dots,M$; there exists a $(2^{mR};2^{mR_1},\dots,2^{mR_M};m)$
code such that $P_e^m<\delta$.
\smallskip
\smallskip

\textit{Proof:} The proof of the theorem for discrete finite-size
alphabets relies on a random coding argument based on the idea of
joint (\textit{strong}) typicality. For the idea of strong
typicality and properties of typical sequences, see \cite{TC91}. The
proof can be outlined as follows. Given $q(x)$ generate a random
channel codebook $\mathcal{C}_c$ with $2^{mR}$ codewords, each of
length $m$, independently from the distribution
$$
q(x^m)=\prod_{k=1}^{m}q(x^m(k)).
$$
and call them $X^m(1),X^m(2),\dots,X^m(2^{mR})$. Also generate $M$
quantization codebooks $\mathcal{C}_j,j=1,\dots,M$, each codebook
$\mathcal{C}_j$ consisting of $2^{mR_j}$ codewords drawn
independently from
$$
p_j(\hat{y}_j^m)=\prod_{k=1}^{m}\sum_{\substack{x\in\mathcal{X}\\y_1\in\mathcal{Y}_1,\dots,y_M\in\mathcal{Y}_M}}q(x)p(y_1,\dots,y_M|x)q_j(\hat{y}_j^m(k)|y_j).
$$
and index them as $\hat{Y}_j^m(1),\hat{Y}_j^m(2),\dots
\hat{Y}_j^m(2^{mR_j})$. Given the message $w$ send the codeword
$X^m(w)$ through the channel. The channel will yield
$Y_1^m,\dots,Y_M^m$. Given the channel output $Y_j^m$ at the $j$'th
quantizer, choose $i_j$ such that $(Y_j^m, \hat{Y}_j^m(i_j))$ are
jointly typical. If there exist no such $i_j$, declare an error. If
the number of codewords in the quantization codebook $2^{mR_j}$ is
greater than $2^{mI(Y_j;\hat{Y}_j)}$, the probability of finding no
such $i_j$ decreases to zero exponentially as $m$ increases. The
probability of failing to find such an index in at least one of the
$M$ quantizers is bounded above by the union bound with the sum of
$M$ exponentially decreasing probabilities in $m$. Given
$\hat{Y}_1^m(i_1),\dots,\hat{Y}_M^m(i_M)$ at the channel decoder,
choose the unique $\hat{w}$ such that
$(X^m(\hat{w}),\hat{Y}_1^m(i_1),\dots,\hat{Y}_M^m(i_M))$ are jointly
typical. The fact that
$(X^m(w),\hat{Y}_1^m(i_1),\dots,\hat{Y}_M^m(i_M))$  will be jointly
typical with high probability can be established by identifying the
Markov chains in the problem and applying Markov Lemma \cite[Lemma
14.8.1]{TC91} repeatedly. Observing that
$(Y_1^m,\dots,Y_M^m,\hat{Y}_1^m,\dots,\hat{Y}_j^m)-Y_{j+1}^m-\hat{Y}_{j+1}^m$
form a Markov chain and recursively applying Markov Lemma, we
conclude that
$(Y_1^m,\dots,Y_M^m,\hat{Y}_1^m(i_1),\dots,\hat{Y}_M^m(i_M))$ are
jointly typical with probability approaching 1 as $m$ increases.
Observing that
$X^m-(Y_1^m,\dots,Y_M^m)-(\hat{Y}_1^m,\dots,\hat{Y}_M^m)$ form
another Markov chain, again by Markov Lemma we have
$(X^m(w),\hat{Y}_1^m(i_1),\dots,\hat{Y}_M^m(i_M))$ jointly typical
with high probability. If there are more than one codewords $X^m$
that are jointly typical with
$(\hat{Y}_1^m(i_1),\dots,\hat{Y}_M^m(i_M))$, we declare an error.
The probability of having more than one such sequence will decrease
exponentially to zero as $m$ increases, if the number of channel
codewords $2^{mR}$ is less than
$2^{mI(X;\hat{Y}_1,\dots,\hat{Y}_M)}$. Hence if
$R<I(X;\hat{Y}_1,\dots,\hat{Y}_M)$ and $R_j>I(Y_j;\hat{Y}_j)$, the
probability of error averaged over all codes decreases to zero as
$m\rightarrow\infty$. This shows the existence of a code that
achieves rates $(R;R_1,\dots,R_M)$ with arbitrarily small
probability of error. The result can be readily extended to
memoryless channels with discrete-time and continuous alphabets by
standard arguments (see \cite[Ch.7]{G91}).\hfill $\square$

\smallskip
\smallskip
\textit{Proof of Lemma~\ref{lem:sec4_4}:} Now we turn to our
original problem: We need to show that it is possible to encode the
observations at the outputs of the MIMO channel at a fixed rate,
while preserving the spatial multiplexing gain of the MIMO channel.
This is a direct consequence of Theorem~II.1: Consider the
conditional probability densities
$$
q_j(\hat{y}_j|y_j)\sim\NC_\CC(y_j,\Delta^2)
$$
for the quantization process. From Theorem~II.1 we know that for any
distribution $p(x)$ on the input space, all rate pairs
$(R;R_1,\dots,R_M)$ are simultaneously achievable if
$$
R_j> I(Y_j;\hat{Y}_j) \qquad j=1,\dots,M \qquad\textrm{and}\qquad R<
I(X;\hat{Y}_1,\dots,\hat{Y}_M)
$$
where now $R_j$ is the encoding rate of the $j$'th stream and $R$ is
the total transmission rate over the MIMO channel. Using
Lemma~\ref{lem:sec4_5}, we have that $I(Y_j;\hat{Y}_j)\leq
\log(1+\frac{P_2}{\Delta^2})$ for any probability distribution
$p(x)$ on the input space. So if we choose
$$
R_j=\log(1+\frac{P_2}{\Delta^2})+\varepsilon\qquad\forall
j=1,\dots,M
$$
for some $\varepsilon>0$, all rates
$$
R\leq I(X;\hat{Y}_1,\dots,\hat{Y}_M)
$$
are achievable on the quantized MIMO channel for any input
distribution $p(x)$. Note that now the channel from $X$ to
$\hat{Y}_1,\dots,\hat{Y}_M$ is given by
$$
\hat{Y}=HX+Z+D
$$
where $D\sim\NC_\CC(0,\Delta^2 I)$. Obviously, this channel has the
same spatial multiplexing gain with the original MIMO channel.
\hfill $\square$

\smallskip
\smallskip
\textit{Proof of Lemma~\ref{lem:sec4_6}:} Consider the case where
the MIMO signals are corrupted by interference of increasing power
$K_I\log M$. In this case, the power received by the destination
nodes is not bounded anymore and increases as $P_2+K_I\log M$ with
increasing $M$. In order to apply the technique employed in the
proof of Lemma~\ref{lem:sec4_4}, one can first normalize the
received signal by multiplying it by $q=\sqrt{\frac{P_2}{P_2+K_I\log
M}}$ and then do the quantization as before. Note that the resultant
{\em scaled} quantized MIMO channel is given by
$$
\hat{Y}=q(HX+Z)+D
$$
where again $D\sim\NC_\CC(0,\Delta^2 I)$ and $Z=(z_k)$ is the
background noise plus interference vector independent of the signal
with uncorrelated entries of power $\EE[z_k^2]\leq N_0+K_I\log M$.
Thus we can apply the result of Lemma~\ref{lem:sec4_3}. Note that
the resultant signal-to-noise-ratio $\SNR\geq\frac{K}{\log M}$ for a
constant $K>0$. Plugging this SNR expression into (\ref{MIMO_gain})
yields $M/\log M$ capacity scaling for the resultant channel. \hfill
$\square$

\section{Largest eigenvalue behaviour of the equalized channel matrix $\tilde{H}$}
\label{app:largest_ev}

In this appendix, we give the proofs of Lemma~\ref{lem:largest_ev}
and Lemma~\ref{lem:dk_scaling}. We  start with
Lemma~\ref{lem:largest_ev}. The proof of the second lemma is given
at the end of the section.

\textit{Proof of Lemma~\ref{lem:largest_ev}:} Let us start by
considering the $2m^{th}$ moment of the spectral norm of $\tilde{H}$
given by  (see \cite[Ch. 5]{horn})
$$
\Vert \tilde{H} \Vert^{2m} = \rho(\tilde{H}^*\tilde{H})^m = \lim_{l
\to \infty} \{\Tr((\tilde{H}^* \tilde{H})^{l})\}^{m/l}.
$$
By dominated convergence theorem and Jensen's inequality, we have
$$
\EE( \Vert \tilde{H} \Vert^{2m} ) \leq \lim_{l \to \infty} \{\EE(
\Tr((\tilde{H}^* \tilde{H})^{l}))\}^{m/l}.
$$
In the subsequent paragraphs, we will prove that the following upper
bound holds with high probability,
\begin{equation} \label{ub}
\EE( \Tr((\tilde{H}^* \tilde{H})^{l})) \leq t_l \, n \,
(K_1^\prime\log n)^{3l}
\end{equation}
where $t_l = \frac{(2l)!}{l! (l+1)!}$ are the Catalan numbers and
$K_1^\prime>0$ is a constant independent of $n$. By Chebyshev's
inequality, this allows to conclude that for any $m$,
$$
\PP(B_{n,\varepsilon}) \leq \frac{\EE(\Vert \tilde{H} \Vert^{2m})}
{n^{m \varepsilon}} \leq \frac{1}{n^{m \varepsilon}} \, \lim_{l \to
\infty} (t_l \, n \, (K_1^\prime\log n)^{3l})^{m/l} \leq \frac{
\left(4 (K_1^\prime\log n)^3\right)^m}{n^{m \varepsilon}},
$$
since $\lim_{l \to \infty} t_l^{1/l}=4$. For any $\varepsilon>0$,
choosing $m$ sufficiently large shows therefore that
$\PP(B_{n,\varepsilon})$ decays polynomially with arbitrary exponent
as $n \to \infty$, which is the result stated in
Lemma~\ref{lem:largest_ev}.

There remains to prove the upperbound in (\ref{ub}). Expanding the
expression gives
\begin{align}\label{expand}
\EE( \Tr((\tilde{H}^*
\tilde{H})^{l}))=\sum_{\substack{i_1,\dots,i_l\in D\setminus V_D\\
k_1,\dots,k_l\in
S}}\EE\left(\overline{\tilde{H}_{i_1k_1}}\tilde{H}_{i_1k_2}\overline{\tilde{H}_{i_2k_2}}\tilde{H}_{i_2k_3}\overline{\tilde{H}_{i_3k_3}}\tilde{H}_{i_3k_4}\dots\overline{\tilde{H}_{i_lk_l}}\tilde{H}_{i_lk_1}\right).
\end{align}
Recall that the random variables $\tilde{H}_{ik}$ are independent
and zero-mean, so the expectation is only non-zero when the terms in
the product form conjugate pairs. Let us consider the case $l=2$ as
an example. We have,
\begin{align}\label{eq:l_1}
\EE( \Tr((\tilde{H}^* \tilde{H})^{2}))&=\sum_{\substack{i_1,i_2\in D\setminus V_D \\ k_1,k_2\in S}}\EE\left(\overline{\tilde{H}_{i_1k_1}}\tilde{H}_{i_1k_2}\overline{\tilde{H}_{i_2k_2}}\tilde{H}_{i_2k_1}\right)\\
&=\sum_{\substack{i_1,i_2\in D\setminus V_D \\ k\in
S}}|\tilde{H}_{i_1k}|^2 |\tilde{H}_{i_2k}|^2+\sum_{\substack{i\in
D\setminus V_D \\ k_1\neq k_2\in
S}}|\tilde{H}_{ik_1}|^2|\tilde{H}_{ik_2}|^2\label{eq:l_2}
\end{align}
since the expectation is non-zero only when either $k_1=k_2=k$ or
$i_1=i_2=i$. Note that we have removed the expectations in
(\ref{eq:l_2}) since $|\tilde{H}_{ik}|^2$ is a deterministic
quantity in our case. The expression can be bounded above by
\begin{align}
\EE( \Tr((\tilde{H}^* \tilde{H})^{2}))\leq\sum_{\substack{i_1,i_2\in
D\setminus V_D \\ k\in S}}|\tilde{H}_{i_1k}|^2
|\tilde{H}_{i_2k}|^2+\sum_{\substack{i\in D\setminus V_D \\ k_1,
k_2\in S}}|\tilde{H}_{ik_1}|^2|\tilde{H}_{ik_2}|^2\label{ub_l2}
\end{align}
where we now doublecount the terms with $i_1=i_2=i$ and $k_1=k_2=k$, that is, the terms of the form $|\tilde{H}_{ik}|^4$.\\

\begin{figure*}[tbp]
\begin{center}
\label{fig:ring_l2}
\input{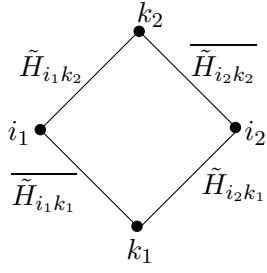}
\end{center}
\caption{The product in Eq.~\ref{eq:l_1} illustrated as a ring.}
\end{figure*}

The non-vanishing terms in the sum in (\ref{eq:l_1}) can also be
determined by the following approach, which generalizes to larger
$l$: let each index be associated to a vertex and each term in the
product in (\ref{eq:l_1}) to an edge between its corresponding
vertices. Note that the resulting graph is in general a ring with
$4$ edges as depicted in Figure 9. A term in the summation in
(\ref{eq:l_1}) is only non-zero if each edge of its corresponding
graph has even multiplicity. Such a graph can be obtained from  the
ring in Figure 9 by merging some of the vertices, thus equating
their corresponding indices. For example, merging the vertices $k_1$
and $k_2$ into a single vertex $k$ gives the graph in Figure 10-a;
on the other hand, merging $i_1$ and $i_2$ into a single vertex $i$
gives Figure 10-b. Note that in the first figure $i_1$, $i_2$ can
take values in $D\setminus V_D$ and $k$ can take values in $S$, thus
the sum of all such terms yields
\begin{figure*}[tbp]
\begin{center}
\label{fig:trees_l2}
\input{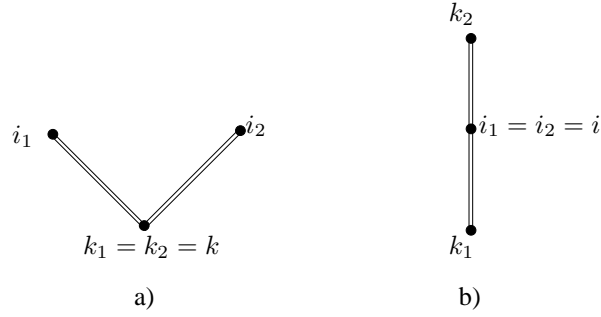}
\end{center}
\caption{Two possible graphs corresponding to the non-zero terms in
(\ref{eq:l_1}).}
\end{figure*}

\begin{equation}\label{l2_term1}
\sum_{\substack{i_1,i_2\in D\setminus V_D \\ k\in
S}}|\tilde{H}_{i_1k}|^2 |\tilde{H}_{i_2k}|^2.
\end{equation}
Similarly, the terms of the form in Figure 10-b sum up to
\begin{equation}\label{l2_term2}
\sum_{\substack{i\in D\setminus V_D \\ k_1, k_2\in
S}}|\tilde{H}_{ik_1}|^2|\tilde{H}_{ik_2}|^2.
\end{equation}
Note that another possible graph composed of edges with even multiplicity can be obtained by further merging the vertices $i_1$ and $i_2$ into a single vertex $i$ in Figure 10-a, or equivalently merging $k_1$ and $k_2$ into $k$ in Figure 10-b. This will result in a graph with only two vertices $k$ and $i$ and a quadruple edge in between which corresponds to terms of the form $|\tilde{H}_{ik}|^4$ with $i\in D\setminus V_D$ and $k\in S$. Note however that such terms have already been considered in both (\ref{l2_term1}) and (\ref{l2_term2}) since we did not exclude the case $i_1=i_2$ in (\ref{l2_term1}) and $k_1=k_2$ in (\ref{l2_term2}). In fact, terms corresponding to any graph with number of vertices less than $3$ are already accounted for in either one of the sums in (\ref{l2_term1}) and (\ref{l2_term2}), or simultaneously in both. Hence, the sum of (\ref{l2_term1}) and (\ref{l2_term2}) is an upper bound for (\ref{eq:l_1}) yielding again (\ref{ub_l2}).\\

\begin{figure*}[tbp]
\begin{center}
\label{fig:ring}
\input{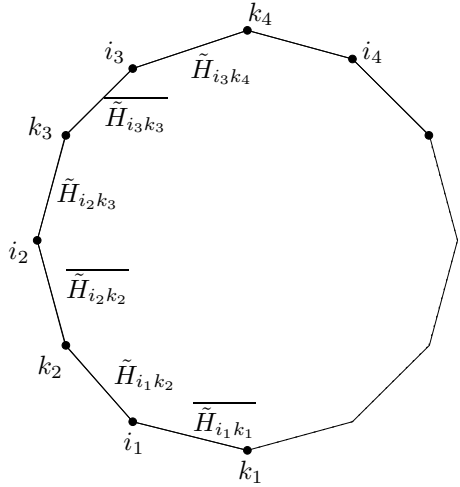}
\end{center}
\caption{The product in Eq.~\ref{expand} illustrated as a ring.}
\end{figure*}

\begin{figure*}[bp]
\begin{center}
\label{fig:trees_l3}
\input{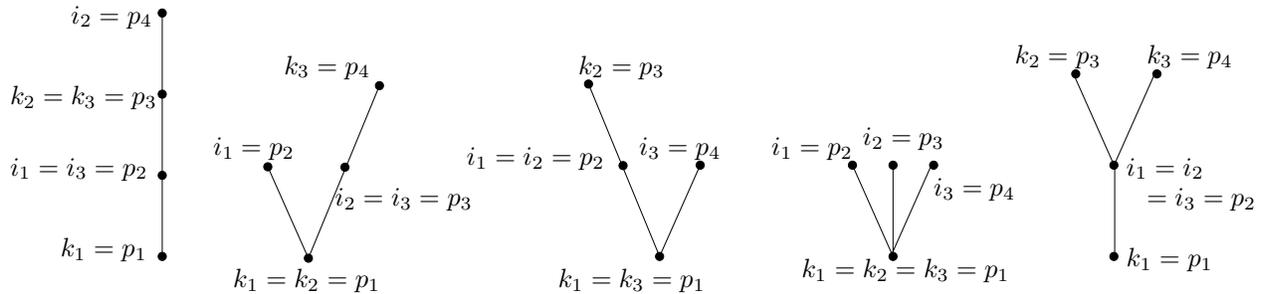}
\end{center}
\caption{Planar rooted planted trees with $3$ branches. Note that
each edge is actually a double edge in our case, although depicted
with a single line in the figure.}
\end{figure*}

In the general case with $l\geq 2$, considering (\ref{expand}) leads
to a larger ring with $2l$ edges, as depicted in Figure 11.
Similarly to the case $l=2$, the non-vanishing terms in
(\ref{expand}) are those that correspond to a graph having only
edges of even multiplicity. Since each edge can have at least double
multiplicity, such graphs can have at most $l$ edges. In turn, a
graph with $l$ edges can have at most $l+1$ vertices which is the
case of a tree. Hence, let us first start by considering such trees;
namely, planar trees with $l$ branches that are rooted (at $k_1$)
and planted, implying that rotating asymmetric trees around the root
results in a new tree. See Figure 12 which depicts the five possible
trees with $l=3$ branches where we relabel the resultant $l+1=4$
vertices as $p_1,\dots,p_4$. In general, the number of different
planar, rooted, planted trees with $l$ branches is given by the
$l$'th Catalan number $t_l$ \cite{Stanley}. In each of these trees,
the $l+1$ vertices $p_1,\dots, p_{l+1}$ take values in either
$D\setminus V_D$ or $S$. Hence, each tree $\mathscr{T}_i^l$
corresponds to a group of non-zero terms,
\begin{equation}\label{tl_terms}
T_i^{l}=\sum_{p_1,\dots, p_{l+1}}f_{\mathscr{T}_{i}^{l}}(p_1,\dots,
p_{l+1}),\qquad i=1,\dots,t_l.
\end{equation}

Note that if a non-vanishing term in (\ref{expand}) corresponds to a
graph with less than $l+1$ vertices, then the corresponding graph
posseses either edges with multiplicity larger than $2$ or cycles,
and this term is already accounted for in either one or more of the
terms in (\ref{tl_terms}). This fact can be observed by noticing
that both edges with large multiplicity as well as cycles can be
untied to get trees with $l$ branches, with some of the $l+1$
indices constrained however to share the same values (see Figure
13). Note that such cases are not excluded in the summations in
(\ref{tl_terms}), thus we have
$$
\EE( \Tr((\tilde{H}^* \tilde{H})^{l}))\leq \sum_{i=1}^{t_l}T_i^{l}.
$$
\begin{figure*}[tbp]
\begin{center}
\label{fig:untie}
\input{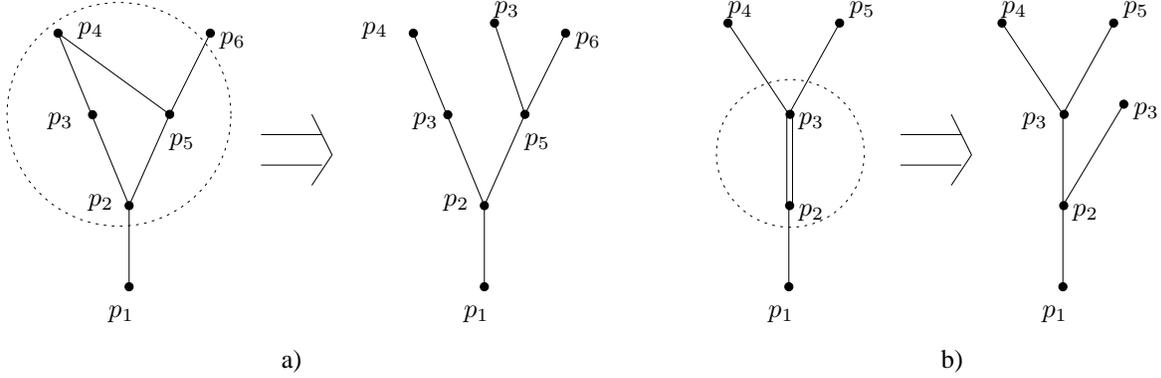}
\end{center}
\caption{The product in Eq.~\ref{expand} illustrated as a ring.}
\end{figure*}

Below we show that
\begin{equation}\label{ub_regular}
T_i^{l}\leq n(K_1^\prime\log n)^{l},\qquad\forall i
\end{equation}
in a regular network and
\begin{equation}\label{ub_random}
T_i^{l}\leq n(K_1^\prime\log n)^{3l},\qquad\forall i
\end{equation}
with high probability in a random network. We first concentrate on
regular networks in order to reveal the proof idea in the simplest
setting. A binning argument then allows to extend the result to
random networks.

{\it a) Regular network:} Recall that in the regular case, the nodes
on the left-half are located at positions $(-k_x+1,k_y)$ and those
on the right half at $(i_x,i_y)$ for
$k_x,k_y,i_x,i_y=1,\ldots,\sqrt{n}$. In this case, the matrix
elements of $\tilde{H}$ are given by
$$
\tilde{H}_{ik} = \frac{e^{j \,
\theta_{ik}}}{((i_x+k_x-1)^2+(i_y-k_y)^2)^{\alpha/4}} \,
\frac{1}{\sqrt{d_{k_x,k_y}}}
$$
and
$$
d_{k_x,k_y} = \sum_{i_x,i_y=1}^{\sqrt{n}}
\frac{1}{((i_x+k_x-1)^2+(i_y-k_y)^2)^{\alpha/2}} .
$$

In the discussion below, we will need an upper bound on the scaling
of $\sum_{i=1}^n |\tilde{H}_{ik}|^2$ and $\sum_{k=1}^n
|\tilde{H}_{ik}|^2$. By Lemma~\ref{lem:dk_scaling}, we have
\begin{align*}
d_{k_x,k_y}\geq K_3^\prime\,k_x^{2-\alpha}
\end{align*}
for a constant $K_3^\prime>0$ independent of $n$ which, in turn,
yields the upper bound
\begin{align}\nonumber
|\tilde{H}_{ik}|^2&\leq\frac{1}{
K_3^\prime}\frac{k_x^{\alpha-2}}{((i_x+k_x-1)^2+(i_y-k_y)^2)^{\alpha/2}}\leq
\frac{1}{K_3^\prime}\frac{1}{(i_x+k_x-1)^2+(i_y-k_y)^2}.
\end{align}
Summing over either $i$ or $k$, and using the upper bound in
Lemma~\ref{lem:dk_scaling} for $\alpha=2$ yields
\begin{equation}\label{ub_logn2}
\sum_{i=1}^n |\tilde{H}_{ik}|^2,\sum_{k=1}^n |\tilde{H}_{ik}|^2\leq
K_1^\prime\log n
\end{equation}
where $K_1^\prime=\frac{K_2^\prime}{K_3^\prime}$ with $K_2^\prime$ and $K_3^\prime$ being the constants appearing in the lemma.\\

Let us first consider the simplest case where the tree is composed
of $l$ height 1 branches and denote it by $\mathscr{T}_1^l$ (see
Figure 14). We have
\begin{figure*}[tbp]
\begin{center}
\label{fig:simpletree}
\input{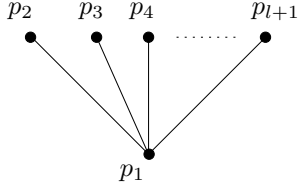}
\end{center}
\caption{A simple tree with $l$ branches.}
\end{figure*}

\begin{align}\nonumber
T_1^{l}=\sum_{p_1,\dots, p_{l+1}=1}^n
f_{\mathscr{T}_{1}^{l}}(p_1,\dots, p_{l+1})&=\sum_{p_1,\dots,
p_{l+1}=1}^n |\tilde{H}_{p_2p_1}|^2\,|\tilde{H}_{p_3p_1}|^2\dots
|\tilde{H}_{p_{l+1}p_1}|^2\\\nonumber
&=\sum_{p_1=1}^n\left(\sum_{p_2=1}^n|\tilde{H}_{p_2 p_1}|^2\right)^l\\
&\leq n(K_1^\prime\log n)^l\label{ub_simple}
\end{align}
which follows from the upper bound (\ref{ub_logn2}).

Now let us consider the general case of an arbitrary tree
$\mathscr{T}_i^{l}$ having $s$ leaves, where $1\leq s\leq l$ (see
Figure 15). Let the indices corresponding to these leaves be
$m_1,\dots,m_{s}$. Let us denote the ``parent'' vertices of these
leaves by $p_{1},\dots,p_{s^\prime}$ and assume that $p_{1}$ is the
common parent vertex of leaves $m_1,\dots, m_{d_1}$; $p_{2}$ is the
common parent vertex of leaves $m_{(d_1+1)},\dots, m_{d_2}$ etc. and
finally $p_{s^\prime}$ is the parent of $m_{(d_t+1)},\dots, m_s$.
The term $T_i^{l}$ corresponding to this tree is given by
\begin{align}
T_i^{l}&=\sum_{\substack{m_1,\dots,m_s\\ p_{1},\dots,p_{(l+1-s)}}}f_{\mathscr{T}_i^{l}}(p_1,\dots, p_{l+1})\nonumber\\
&=\sum_{p_1,\dots,p_{(l+1-s)}=1}^nf_{\mathscr{T}_{i^\prime}^{l-s}}(p_1,\dots,p_{(l+1-s)})\nonumber\\
&\times\sum_{m_1,\dots,m_{s}=1}^n|\tilde{h}_{m_1 p_{1}}|^2\dots
|\tilde{H}_{m_{d_1} p_1}|^2\,|\tilde{H}_{m_{(d_1+1)} p_{2}}|^2\dots
|\tilde{H}_{m_{d_2} p_{2}}|^2\,\,\dots\,\,\, |\tilde{H}_{m_{(d_t+1)}
p_{s^\prime}}|^2\dots |\tilde{H}_{m_s
p_{s^\prime}}|^2\label{eq:leaves}\\\label{eq:leaves2} &\leq
T_{i^\prime}^{l-s}(K_1^\prime\log n)^s
\end{align}
where $T_{i^\prime}^{l-s}$ corresponds to a smaller (and shorter)
tree $\mathscr{T}_{i^\prime}^{(l-s)}$ with $l-s$
branches\footnote{Note that the term corresponding to a leaf $m$ can
be either $|\tilde{H}_{m p}|^2$ or $|\tilde{H}_{pm}|^2$ depending on
whether the height of the leaf is even or odd. However, in
(\ref{eq:leaves}), we ignore this issue in order to simplify the
notation since the upper bound (\ref{eq:leaves2}) applies in both
cases.}. The argument above decreases the height of the tree by 1,
hence can be applied recursively to get a simple tree composed only
of height $1$ branches in which case the upper bound in
(\ref{ub_simple}) applies. Thus, given $\mathscr{T}_i^{l}$ let $h$
be the number of recursions to get a simple tree and $s_1,\dots,s_h$
denote the number of leaves in the trees observed at each step of
the recursion. We have
\begin{align*}
T_i^{l}&\leq (K_1^\prime\log n)^{s_1}(K_1^\prime\log n)^{s_2}\dots(K_1^\prime\log n)^{s_h}T_1^{l-s_1\dots-s_h}\\
&\leq n(K_1^\prime\log n)^{l}
\end{align*}
since $T_1^{l-s_1\dots-s_h}\leq n(K_1^\prime\log n)^{l-s_1\dots-s_h}$ by (\ref{ub_simple}). Thus, (\ref{ub_regular}) follows.\\

\begin{figure*}[tbp]
\begin{center}
\label{fig:leaves}
\input{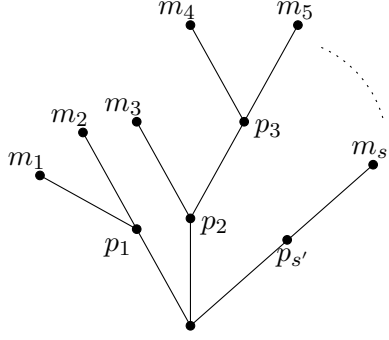}
\end{center}
\caption{A tree with leaves $m_1,m_2,\dots,m_s$.}
\end{figure*}

{\it b) Random network:} We denote the locations of the nodes to the
left of the cut  by $a_k=(-a^x_{k},a^y_{k})$ where $a^x_{k}$ is the
$x$-coordinate and $a^y_{k}$ is the $y$-coordinate of node $k\in S$
and those to the right of the cut are similarly denoted by
$b_{i}=(b^x_{i},b^y_{i})$ for $i\in D\setminus V_D$. In this case,
the matrix elements of $\tilde{H}$ are given by
$$
\tilde{H}_{ik} = \frac{e^{j \,
\theta_{ik}}}{((b^x_{i}+a^x_{k})^2+(b^y_{i}-a^y_{k})^2)^{\alpha/4}}
\, \frac{1}{\sqrt{d_{k}}}
$$
and
$$
d_{k} = \sum_{i\in D\setminus V_D}
\frac{1}{((b^x_{i}+a^x_{k})^2+(b^y_{i}-a^y_{k})^2)^{\alpha/2}}.
$$
In parallel to the regular case, we will need an upper bound on
$\sum_{i\in D\setminus V_D} |\tilde{H}_{ik}|^2$ and $\sum_{k\in S}
|\tilde{H}_{ik}|^2$. The upper bound can be obtained in two steps by
first showing that
\begin{equation}\label{lb}
d_{k}\geq K_3^\prime\frac{(a^x_{k})^{2-\alpha}}{\log n}
\end{equation}
with high probability for a constant $K_3^\prime>0$ independent of
$n$,  which leads to
\begin{equation}\label{ub_h}
|H_{ik}|^2 \leq \frac{1}{K_3^\prime}\log
n\,\frac{(a^x_{k})^{\alpha-2}}{((b^x_{i}+a^x_{k})^2+(b^y_{i}-a^x_{k})^2)^{\alpha/2}}\leq\frac{1}{K_3^\prime}\log
n\, \frac{1}{(b^x_{i}+a^x_{k})^2+(b^y_{i}-a^y_{k})^2}
\end{equation}
for all $i,k$. This, in turn yields
\begin{equation}\label{ub_random2}
\sum_{k\in S} |\tilde{H}_{ik}|^2,\,\sum_{i\in D\setminus V_D}
|\tilde{H}_{ik}|^2\leq K_1^\prime(\log n)^3
\end{equation}
with high probability for another constant $K_1^\prime>0$ independent of $n$. Recalling the leaf removal argument discussed for regular networks immediately leads to (\ref{ub_random}).\\

Both the lower bound in (\ref{lb}) and the upper bound in
(\ref{ub_random2}) regarding random networks can be proved using

binning arguments that provide the connection to regular networks.
In order to prove the lower bound, we consider Part (b) of
Lemma~\ref{lem:logn}, while the upper bound (\ref{ub_random2}) is
proved using Part (a) of the same lemma.

Let us first consider dividing the right-half network into
squarelets of area $2\log n$. Given a left-hand side node $k$
located at $(-a^x_{k},a^y_{k})$, let us move the nodes inside each
right-hand side squarelet onto the squarelet vertex that is farthest
to $k$. Since this displacement can only increase the Euclidean
distance between the nodes involved, and since by Part (b) of
Lemma~\ref{lem:logn}, we know that there is at least one node inside
each squarelet, we have
\begin{align*}
d_{k} &= \sum_{i\in D\setminus V_D} \frac{1}{((b^x_{i}+a^x_{k})^2+(b^y_{i}-a^y_{k})^2)^{\alpha/2}}\\
&\geq\sum_{i_x,i_y=1}^{\sqrt{n/2\log n}} \frac{1}{((i_x\sqrt{2\log n}+a^x_{k})^2+(i_y\sqrt{2\log n}-a^x_{k})^2)^{\alpha/2}}\\
&\geq K_3^\prime \frac{(a^x_{k})^{2-\alpha}}{2\log n}
\end{align*}
by using the lower bound in Lemma~\ref{lem:dk_scaling}.

Now having (\ref{ub_h}) in hand, in order to show
(\ref{ub_random2}), we divide the network into $n$ squarelets of
area 1. By Part (a) of Lemma~\ref{lem:logn}, there are at most $\log
n$ nodes inside each squarelet. Considering the argument in
Section~\ref{sec:extended} and the displacement of the nodes as
illustrated in Figure 7 yields a regular network with at most $2\log
n$ nodes at each vertex in the right-half network,
\begin{align*}
\sum_{i\in D\setminus V_D}|\tilde{H}_{ik}|^2&\leq \frac{2}{K_3^\prime}\log n \sum_{i\in D\setminus V_D}\frac{1}{(b^x_{i}+a^x_{k})^2+(b^y_{i}-a^y_{k})^2}\\
&\leq \frac{4}{K_3^\prime}(\log n)^2\sum_{i_x,i_y=1}^{\sqrt{n}}\frac{1}{(i_x+k_x)^2+(i_y-k_y)^2}\\
&\leq 4K_1^\prime(\log n)^3.
\end{align*}
by employing the upper bound in Lemma~\ref{lem:dk_scaling} for
$\alpha=2$. The same bound follows similarly for $\sum_{k\in
S}|\tilde{H}_{ik}|^2$, thus the desired result in
(\ref{ub_random2}). \hfill $\square$

{\it Proof of Lemma~\ref{lem:dk_scaling}:} Both the lower and upper
bound for $d_{k_x,k_y}$ can be obtained by straightforward
manipulations. Recall that
$$
d_{k_x,k_y} = \sum_{i_x,i_y=1}^{\sqrt{n}}
\frac{1}{((i_x+k_x-1)^2+(i_y-k_y)^2)^{\alpha/2}} .
$$
The upper bound can be obtained as follows:
\begin{align*}
d_{k_x,k_y}&=\sum_{y=1-k_y}^{\sqrt{n}-k_y}\sum_{x=k_x}^{k_x+\sqrt{n}-1}
\frac{1}{(x^2+y^2)^{\alpha/2}}
\leq \sum_{y=1-k_y}^{\sqrt{n}-k_y}\left( \frac{1}{(k_x^2+y^2)^{\alpha/2}}+\int_{k_x}^{k_x+\sqrt{n}-1}\frac{1}{(x^2+y^2)^{\alpha/2}}\,dx\right)\\
&\leq k_x^{-\alpha}\,+\,\int_{k_x}^{k_x+\sqrt{n}-1}\frac{1}{x^{\alpha}}dx\,+\,\int_{1-k_y}^{\sqrt{n}-k_y} \frac{1}{(k_x^2+y^2)^{\alpha/2}}\,dy\\
&\quad +\int_{1-k_y}^{\sqrt{n}-k_y}\int_{k_x}^{k_x+\sqrt{n}-1}\frac{1}{(x^2+y^2)^{\alpha/2}} \,dx\,dy\\
& \leq
k_x^{-\alpha}\,+\,(1+\pi)k_x^{1-\alpha}\,+\,\int_{-\pi/2}^{\pi/2}\int_{k_x}^{3\sqrt{n}}\frac{1}{r^\alpha}
r\,dr\,d\theta
\end{align*}
So
\begin{align}
d_{k_x,k_y}&=\left\{
\begin{array}{ll}
k_x^{-\alpha}\,+\,(1+\pi)k_x^{1-\alpha}\,+\pi\log r\Big|_{k_x}^{3\sqrt{n}} & \text{if } \alpha=2,\\
k_x^{-\alpha}\,+\,(1+\pi)k_x^{1-\alpha}\,+\frac{\pi}{(2-\alpha)}r^{2-\alpha}\Big|_{k_x}^{3\sqrt{n}}
& \text{if } \alpha>2,
\end{array} \right. \label{appIIItemp1}\\
&\leq\left\{
\begin{array}{ll}
K_2^\prime\,\log n & \text{if } \alpha=2,\\
K_2^\prime\,k_x^{2-\alpha} & \text{if } \alpha>2,
\end{array} \right. \nonumber
\end{align}
for a constant $K_2^\prime>0$ independent of $n$, since the dominating terms in (\ref{appIIItemp1}) are the third ones.\\

The lower bound follows similarly:
\begin{align*}
d_{k_x,k_y}&=\sum_{y=1-k_y}^{\sqrt{n}-k_y}\sum_{x=k_x}^{k_x+\sqrt{n}-1}
\frac{1}{(x^2+y^2)^{\alpha/2}}
\geq\sum_{y=1-k_y}^{\sqrt{n}-k_y}\int_{k_x}^{k_x+\sqrt{n}-1}\frac{1}{(x^2+y^2)^{\alpha/2}}\,dx\\
&\geq\int_{1-k_y}^{\sqrt{n}-k_y}\int_{k_x}^{k_x+\sqrt{n}-1}\frac{1}{(x^2+y^2)^{\alpha/2}}\,dx\,dy-\int_{k_x}^{k_x+\sqrt{n}-1}\frac{1}{x^\alpha}\,dx\\
&\geq\int_{0}^{\sqrt{n}}\int_{k_x}^{k_x+\sqrt{n}-1}\frac{1}{(x^2+y^2)^{\alpha/2}}\,dx\,dy +\frac{x^{1-\alpha}}{\alpha-1}\Big|_{k_x}^{k_x+\sqrt{n}-1}\\
&\geq\int_{0}^{\arctan(1/2)}\int_{\sqrt{2}k_x}^{k_x+\sqrt{n}-1}\frac{1}{r^{\alpha}}\,r\,dr\,d\theta
+\frac{x^{1-\alpha}}{\alpha-1}\Big|_{k_x}^{k_x+\sqrt{n}-1}
\end{align*}
So for all $\alpha\geq 2$, we have
\begin{align}
d_{k_x,k_y}&=\left\{
\begin{array}{ll}
\arctan(\frac{1}{2})\,\log r\Big|_{\sqrt{2}k_x}^{k_x+\sqrt{n}-1}+\frac{1}{\alpha-1}x^{1-\alpha}\Big|_{k_x}^{k_x+\sqrt{n}-1} & \text{if } \alpha=2, \label{appIIItemp2}\\
\arctan(\frac{1}{2})\frac{1}{2-\alpha}r^{2-\alpha}\Big|_{\sqrt{2}k_x}^{k_x+\sqrt{n}-1}+\frac{1}{\alpha-1}x^{1-\alpha}\Big|_{k_x}^{k_x+\sqrt{n}-1}
& \text{if } \alpha>2,
\end{array} \right.\\
&\geq K_3^\prime k_x^{2-\alpha} \nonumber
\end{align}
where $K_3^\prime>0$ is a constant independent of $n$, since the dominating terms in (\ref{appIIItemp2}) are the first ones. This concludes the proof of the lemma. \hfill $\square$\\


\end{document}